\documentclass{emulateapj}

\def\xmm{{XMM-{\it Newton\/}}}
\def\fermi{{{\it Fermi}-LAT}}

\newcommand{\fhl}{2FHL~J1703.4--4145}

\shorttitle{$\gamma$-ray emission revealed at the western edge of SNR G~344.7-0.1}

\shortauthors{Eagle et al.}

\PassOptionsToPackage{colorlinks,linkcolor={blue},citecolor={blue},urlcolor={red}}{hyperref}
\usepackage{hyperref}
\usepackage{natbib}
\usepackage{float}
\usepackage{color}
\usepackage{graphicx}
\usepackage{gensymb}
\usepackage{enumitem}
\usepackage{multirow,helvet}

\begin{document}


%\title{Unveiling Extreme Galactic Accelerators: Discovery of shock-cloud interaction on Western edge of Vela SNR}

\title{{\large{$\gamma$}}-ray emission revealed by the {\it Fermi}-LAT at the western edge of SNR~G344.7--0.1}

\author{J. Eagle\altaffilmark{1, 2}, S. Marchesi\altaffilmark{1}, D. Castro\altaffilmark{2}, M. Ajello\altaffilmark{1}, A. Vendrasco\altaffilmark{1}}
%\affil{Astronomy Department, University of California}
%\today

\altaffiltext{1}{Department of Physics \& Astronomy, Clemson University, Clemson, SC 29634, USA}
\altaffiltext{2}{Harvard-Smithsonian Center for Astrophysics, Cambridge, MA 02138, USA}
%\altaffiltext{3}{Universidad de Buenos Aires,Facultad de Cs Exactas y Naturales, Buenos Aires, Argentina}
%\altaffiltext{4}{CONICET - Universidad de Buenos Aires, Instituto de Astronomía y Física del Espacio (IAFE), Buenos Aires, Argentina}
%\altaffiltext{5}{IRAP/Observatoire Midi-Pyrénées, France}

%\slugcomment{Credit to Dubner and her grad student for their participation as well.}

\begin{abstract}
We report on the investigation of a very high energy (VHE), Galactic $\gamma$-ray source recently discovered at $>$50\,GeV using the Large Area Telescope (LAT) on board the {\it Fermi} Gamma-Ray Space Telescope (\textit{Fermi}). This object, 2FHL~J1703.4--4145, displays a very hard $>$50\,GeV spectrum with a photon index $\Gamma_{\gamma}\sim1.2$ in the 2FHL catalog and, as such, is one of the most extreme sources in the 2FHL sub-sample of Galactic objects. A detailed analysis of the available multi-wavelength data shows that this source is located on the western edge of the supernova remnant (SNR) G344.7--0.1, along with extended TeV source, HESS~J1702--420. The observations and the spectral energy distribution modeling support a scenario where this $\gamma$-ray source is the byproduct of the interaction between the SNR shock and the dense surrounding medium, with escaping cosmic rays (CRs) diffusing into the dense environment and interacting with a large local cloud, generating the observed TeV emission. If confirmed, an interaction between the SNR CRs and a nearby cloud would make 2FHL~J1703.4--4145 another promising candidate for efficient particle acceleration of the 2FHL Galactic sample, following the first candidate from our previous investigation of a likely shock-cloud interaction occurring on the West edge of the Vela SNR. 
\end{abstract}
 %H.E.S.S., CR diffusion, CR escape, local CR accelerators, etc.

\section{Introduction}
%\large update findings in IR section to include whatever the 24um paper reports as interesting for us. Also the 1997 OH report. add citations. Lastly, update the parameter table and image of the final X-ray model.--Jordan 09/27/2019}
%\textcolor{blue}{The introduction can be similar to the one of your previous paper, but it cannot be exactly the same, so you need to do some rephrasing. I think some parts can be slightly shortened now, and you can instead refer to your previous paper adding a few more details, particularly on the approach adopted to determine that the 2FHL source is a likely shock-cloud interaction.}
%\todo[inline]{New X-ray plot. Cut off at 0.9keV}
Efficient particle accelerators responsible for Galactic cosmic rays (CRs) are abundant in the Milky Way Galaxy, whose interactions with the ambient medium and photon fields produce energetic $\gamma$-rays. Therefore, $\gamma$-rays provide an excellent way to probe non-thermal astrophysical processes. Relativistic electrons (i.e. leptons) can produce $\gamma$-rays by non-thermal bremsstrahlung from Coulomb interactions with ions or by inverse Compton scattering (ICS) on ambient photon fields, whereas protons and heavier nuclei (i.e. hadrons) produce $\gamma$-ray emission via hadronic collisions with ambient material generating pions that then decay quickly to $\gamma$-rays. Studies of the non-thermal Galactic source population are essential to understand how and where the bulk of cosmic rays are being accelerated and to understand the mechanisms underlying very high energy (VHE, E$>$50\,GeV) emitters \citep{renaud2009,karg2013}.

Several deep sky observations have been performed to study the Galactic plane in the TeV $\gamma$-ray energy band with facilities like the H.E.S.S., MAGIC, and VERITAS ground-based Cherenkov telescopes \citep{hess2003,magic2005,ver2006,antonelli2009}. These surveys revealed that the Galactic plane is rich with TeV $\gamma$-ray emission from systems leftover from supernova explosions such as pulsar wind nebulae (PWNe) and supernova remnants \citep[SNRs,][]{funk2005, aha2006, carrigan2013, ong2014}. Recently, the Pass 8 \citep{atwood2013} event level reconstruction and analysis has enabled the {\it Fermi-}Large Area Telescope (LAT) to achieve similar sensitivity and coverage to the aforementioned facilities at energies above 50\,GeV, reaching an average sensitivity in the plane of $\sim2\%$ of the Crab flux \citep[only slightly less sensitive than H.E.S.S. in this energy band, see][]{hess2018} with a localization accuracy better than 3$^\prime$ for most sources \citep{ackermann2015}. 

%\textit{Fermi}'s main advantage is that it has surveyed the entire sky, and hence the Galactic plane, with uniform sensitivity and coverage whereas other telescopes are limited to detection from ground-based locations (e.g. H.E.S.S., VERITAS, and MAGIC are all ground-based) and are restricted from viewing the entire Galactic plane with uniform sensitivity. As a result, the \textit{Fermi}-LAT \citep{atwood2013, pass82017} has detected several new Galactic sources, some of which display very hard spectra above 50\,GeV, which is a sign of efficient particle acceleration and (or) effective particle and energy dissipation processes. Understanding the properties of the VHE Galactic source population is crucial in order to identify the locations and mechanisms for Galactic cosmic ray acceleration. 

One breakthrough possible with Pass 8 has been the survey of the entire sky at $>$50\,GeV reported in the 2FHL catalog \citep[][]{ackermann2015}, which is comprised of 360 sources. Of these objects, 103  are detected in the Galactic plane ($\mid b\mid$ $< 10\degree$): 38 of these have been associated with Galactic objects as their counterparts, 42 are associated with blazars, and 23 are unassociated. None of the 23 unassociated sources have the radio or optical properties of blazars, though it may still be possible to have blazars present in the sub-sample. A second selection criterion is applied to classify sources as Galactic in origin by considering the hardness of the $\gamma$-ray spectrum at $\>$50\,GeV, since at these energies blazars generally exhibit a soft spectrum with an average photon index of $\Gamma\sim3.4$ in the 2FHL catalog. This is because the energy range is above the inverse Compton (IC) peak of their spectral energy distribution (SED). 

%This is a result of the combination of the spectral shape of the energy distribution of the accelerated particles and the absorption due to the extragalactic background light \citep{dominguez2011}, which results in an exponentially cut-off photon spectrum. Only $\sim$4\,\% of the 2FHL blazars display a power law photon index $\Gamma<1.8$. 

Among the 23 unidentified 2FHL objects located in the Galactic plane, 12 have $\Gamma<1.8$, and hence the number of contaminant blazars in this hard-spectrum sub-sample is expected to be $<1$. Thus, this sub-sample should be comprised of newly detected hard-spectrum Galactic objects. In \citet{eagle2019}, we report the first findings from the sub-sample on 2FHL~J0826.1--4500 which is found to be a probable shock-cloud interaction on the western edge of the Vela SNR (hereafter referred to as Paper Ia). In this work, we focus on another object in our sample, 2FHL~J1703.4--4145, which is similarly located at the western rim of a supernova remnant, SNR~G344.7--0.1. 

\subsection{SNR~G344.7--0.1}
The discovery of SNR~G344.7--0.1 was first reported by \citet{caswell1975} with observations at 408 and 5000\,MHz. Since then, it has been established that the SNR is relatively young to middle-aged with an age estimated to be $\tau\sim3,000$\,yr and is roughly 8$^\prime$ in diameter in the radio with the brightest radio emission seen to be in the northern, central, and western regions of the remnant \citep{dubner1993,whiteoak1996}. \citet{dubner1993} determined the distance of the SNR to be $d\approx14$\,kpc and in \citet{giacani2011} it is estimated to be 6.3\,kpc from HI absorption measurements. \citet{yama2011} argues that 6.3\,kpc is far too close given the high absorbing column density, $N_H$, measured in the direction of the SNR in the X-ray and must be at least as far as the Galactic tangent point of 8\,kpc. Considering the uncertainty of the SNR distance, we adopt the far and near distances of 14 and 6.3\,kpc, respectively, throughout this paper.

The location of the SNR is coincident with a high density region with plumes of CO, neutral hydrogen, and dust (noticed by abundant IR emission at $24$\,$\mu$m) surrounding the SNR radio shell \citep{giacani2011}. Because of the local medium's tumultuous environment in this region, it has been suggested that the increased radio surface brightness and observed IR emission to the West of the remnant indicates that it is interacting with the ambient interstellar medium \citep[ISM,][]{combi2010, giacani2011,chawner2019}. It would seem plausible then for the SNR to be the result of a core collapse (CC) supernova (SN) explosion as these cloudy, dense environments indicate a massive star exploded not far from its original birthplace. However, the stellar progenitor of this system is yet to be firmly identified because no compact remnant has been identified within the SNR to date which challenges the CC origin theory \citep{combi2010, giacani2011, yama2011}. \citet{combi2010} attempted to tie the unresolved compact X-ray source detected by both \xmm\ and {\it{Chandra}}, CXOU~J170357.8--414302, to the SNR due to its positional overlap with the center of the remnant. However, the characteristics of the potential central compact object (CCO) in the optical and infrared fit more with a K0 dwarf star in the foreground. The CCO also requires a drastically different absorbing column density than what is estimated for the SNR. 

Other attempts to classify SNR~G344.7--0.1 as a CC SN include \citet{chang2008} where an X-ray study was performed to try to identify the extended dark TeV source, HESS~J1702--420 as the PWN around the pulsar, PSR~J1702--4128, that could be a displaced progenitor of the SNR. If this could be confirmed, it would be a persuasive argument that the SNR descended from a CC SN explosion, however, results from \citet{chang2008} are inconclusive. It is also worth mentioning that the estimated age of PSR~J1702--4128 is 55\,kyr with a distance of $d\approx5.2$\,kpc, whereas the SNR is $\sim 3,000$\,yr old\footnote{This estimate is based on the ionization timescale $\tau/n_e$, taking $n_e\sim1$\,cm$^{-3}$ of the observed X-ray emission.} \citep{giacani2011} and is at least 6.3\,kpc away \citep[][]{giacani2011,yama2011}. Therefore, this scenario is unlikely.

\citet{yama2011} suggests this SNR is more likely the result of a type Ia SN, based on strong Fe K-shell emission detected in the X-rays (6-7\,keV) likely emitted from the SN ejecta. Fe emission is frequently observed among type Ia SNRs, while it would be unusual to be found in CC SNRs \citep{yama2011}. For this reason, we could indeed be viewing a unique case of a type Ia SNR interacting with a dense, inhomogeneous surrounding medium.

This paper describes the analysis of existing X-ray observations of the SNR in the region coincident with 2FHL~J1703.4--4145, as well as archival multi-wavelength data in the region where 2FHL~1703.4--4145 is located, how the dark TeV source HESS~J1702--420 may play a role in this region, and the modeling of the broadband spectral energy distribution to better understand the origin of the observed $\gamma$-ray emission. 

\subsection{HESS~J1702-420}
Extended and yet unidentified source, HESS~J1702--420, located at (R.A., Dec.)$=(255.63\degree, -42.07\degree) \pm0.05\degree$ in J2000, is a likely TeV counterpart of 2FHL~J1703.4--4145 due to their positional coincidence with SNR~G344.7--0.1 and compatible $\gamma$-ray spectral energy distributions \citep[see Figure \ref{fig:sed_gamma_map} and][]{hess2018}.

As previously mentioned, HESS~J1702--420 was briefly investigated as a possible displaced PWN to an unseen pulsar or to a detected but displaced pulsar \citep[e.g. PSR~J1702--4128 in][]{chang2008} associated to the SNR, however, the results remain inconclusive and the nature of the TeV source is still unknown. %The nature for the association of HESS~J1702--420 with SNR~G344.7--0.1 and 2FHL~J1703.4--4145 is still unclear. 

A second possibility explaining the extended TeV emission with its peak seen unusually far from the SNR could be runaway CR diffusion that is hitting a nearby molecular cloud and illuminating it. This hypothesis seems plausible based on observationally constrained Monte Carlo simulations \citep{cui2016}. It would also explain the GeV $\gamma$-ray emission observed with the \fermi\ as mainly the result of leaked GeV CRs from a shock-cloud collision with a molecular cloud core \citep{cui2019}. In summary, the displaced TeV emission from an SNR can be described by runaway TeV CRs released early on from the SNR that diffuse rapidly and travel distances of 10-100\,pc. Meanwhile, the GeV CRs escape and diffuse slower and will be mostly concentrated at the shock-cloud boundary. 

If HESS~J1702--420 is indeed the TeV counterpart of 2FHL~J1703.4--4145 as the SED suggests (see Figure~\ref{fig:sed_gamma_map}), and both are associated to SNR~G344.7--0.1, a possible scenario is one where the SNR is interacting with a cloud at the western boundary, explaining the concentration of GeV emission at the location of 2FHL~J1703.4--4145. Subsequently, HESS~J1702--420 could be explained by a dense cloud located at the TeV source position being penetrated and illuminated by runaway TeV CRs that escaped from the SNR at an earlier time.

On the contrary, \citet{lau2019} considers the same scenario in which CRs are accelerated by the SNR, diffuse into the ISM, and thus generate HESS~J1702--420 as the CRs interact with a molecular cloud here. The authors find CR escape and diffusion into a nearby cloud to be an unlikely explanation for the TeV source, determined by using a diffusion time relationship from \citet{ginz1965}. The results suggest the diffusion time for escaped CRs to be much greater than the SNR age. However, we point out that the estimates by \citet{lau2019} are far too large, both the derived diffusion times and SNR ages. \citet{lau2019} use the diffusion coefficient equation from \citet{gabici2005} however, this derivation only considers existing clouds embedded in the diffuse Galactic cosmic ray population and does not account for a local CR accelerator like an SNR. In fact, \citet{gabici2009} derive the diffusion coefficient considering both the diffuse Galactic CR flux and the contribution of CRs from a local CR accelerator, specifically of an SNR. The coefficients are more than a magnitude in difference when accounting for SNR-accelerated CRs, therefore the improved estimate for the diffusion coefficient is
\begin{equation}
    D(E) = 10^{28}\big(\frac{E}{10\text{\,GeV}}\big)^{0.5} \text{cm$^{2}$ s$^{-1}$}
\label{eq:D}\end{equation}
where $E$ is the CR energy. We also invoke the relationship from \citet{ginz1965} of $\tau_{diff}=\frac{d^2}{6D(E)}$ where $d$ is the distance to the TeV emission peak from the SNR, corresponding to $d\sim120$\,pc and $d\sim55$\,pc in the 14\,kpc and 6.3\,kpc SNR distance scenarios, respectively. We use the same CR energy input as \citet{lau2019} $E_{CR}=2$\,TeV which would generate $\gamma$-rays with energy $\sim200$\,GeV, close to the detection threshold of H.E.S.S. Using the diffusion coefficient in (\ref{eq:D}) instead, the estimated diffusion times become $\tau_{diff}\approx$\,5.2\,kyr and $\tau_{diff}\approx$\,1.1\,kyr for the estimated SNR distances of 14\,kpc and 6.3\,kpc, respectively \citep[compared to the estimates of $\tau_{diff}\approx$\,100\,kyr and $\tau_{diff}\approx$\,34\,kyr,][] {lau2019}. 

%Based on the estimated distance of 14 kpc to G344.7−0.1 (Dubner et al. 1993), the separation between G344.7−0.1 and the TeV peakof HESS J1702–420 is d ∼ 120 pc, and the corresponding diffusiontime is calculated to be τ diff ∼ 100 kyr. The radius of G344.7−0.1 atthe distance of 14 kpc is ∼16pc. Employing the model of an SNR inSedov phase from Chevalier (1974), the age of the remnant would be ∼38 kyr.

We also re-estimated the SNR ages using the Sedov-Taylor phase of an SNR \citep{taylor1950,sedov1959},
\begin{equation}
    \big(\frac{t}{100\text{\,yr}}\big)^{\frac{2}{5}} = \frac{R_{sh}}{2.3\text{\,pc}} \big(\frac{E}{10^{51}\text{\,ergs}}\big)^{\frac{1}{5}}\big(\frac{\rho_{0}}{10^{-24}\text{\,g cm$^{-3}$}}\big)^{-\frac{1}{5}}
\end{equation}
where $R_{sh}$ is the shock radius, $E$ is the SN explosion energy, and $\rho_{0}$ is the ambient density. At a distance of 14\,kpc, $R_{sh}\sim16$\,pc and at 6.3\,kpc, $R_{sh}\sim7$\,pc. We assume a $\rho_{0}$ corresponding to an ambient particle density of $n_{0}=1$\,cm$^{-3}$ and $E_{SN}=10^{51}$\,ergs. We find that at a distance of 14\,kpc, the SNR age would be $\sim 10$\,kyr and at 6.3\,kpc, the SNR age would be $\sim1.2$\,kyr, significantly lower than the estimates found in \citet{lau2019}.
%Based on the authors' findings, the TeV emission can be best explained by a hadronic population from a local CR accelerator that has yet to be detected, such as a young SNR. It could also be explained by a yet undetected associated pulsar to the SNR itself. 

A third explanation of HESS~J1702--420 in association with the SNR could be that it is a TeV halo from a displaced pulsar and PWN that is yet to be discovered. TeV halos are spatially extended, non-thermal high-energy $\gamma$-ray emission surrounding a PWN. TeV halos are much larger in extension than PWN but are close enough to the central pulsar that this region is still dominated by pulsar activity and cosmic ray diffusion \citep{halo2019}. The $\gamma$-ray emission is produced by escaped $\sim$ 10\,TeV electrons and positrons from the termination shock of the PWN, scattering off of the interstellar radiation field. TeV halos are observed to have a hard spectrum in this regime with a photon index $\sim2.2$ \citep{halo2019}. HESS~J1702-420 is best fit with a power law index of $2.1\pm0.1$, in good agreement with TeV halo observations. If this scenario is confirmed, this would indicate SNR~G344.7--0.1 as a CC supernova remnant with a clear Fe-K emission line present in the X-ray spectrum, which would be a peculiar finding for a CC SN \citep{yama2011}.

This paper is organized as follows: in Sections \ref{sec:gdata} and \ref{sec:xdata} we discuss the source selection and the XMM-\textit{Newton} data reduction and analysis. A further multi-wavelength characterization of the source is presented in Section \ref{sec:multi}. Section \ref{sec:discuss} explores the depicted scenario through the SED modeling, and Section \ref{sec:conclude} summarizes our results.

\section{Source selection}\label{sec:gdata}
\begin{figure*}[htbp]
\begin{minipage}[b]{.5\textwidth}
%\hspace{3cm}
\centering
\includegraphics[width=0.8\linewidth]{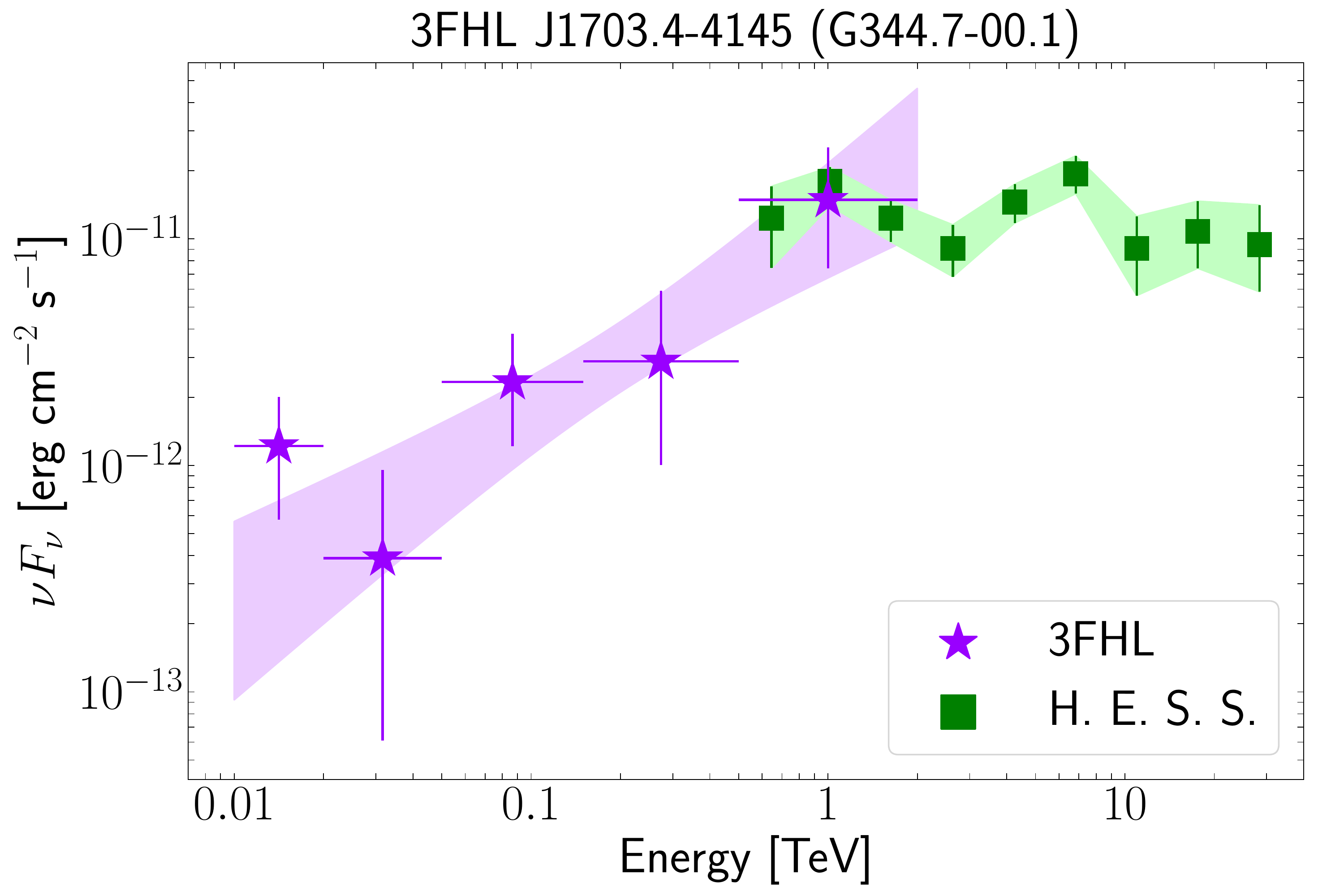}
\end{minipage}
\begin{minipage}[b]{.5\textwidth}
\hspace{-.4cm}
\centering
\includegraphics[width=0.8\linewidth]{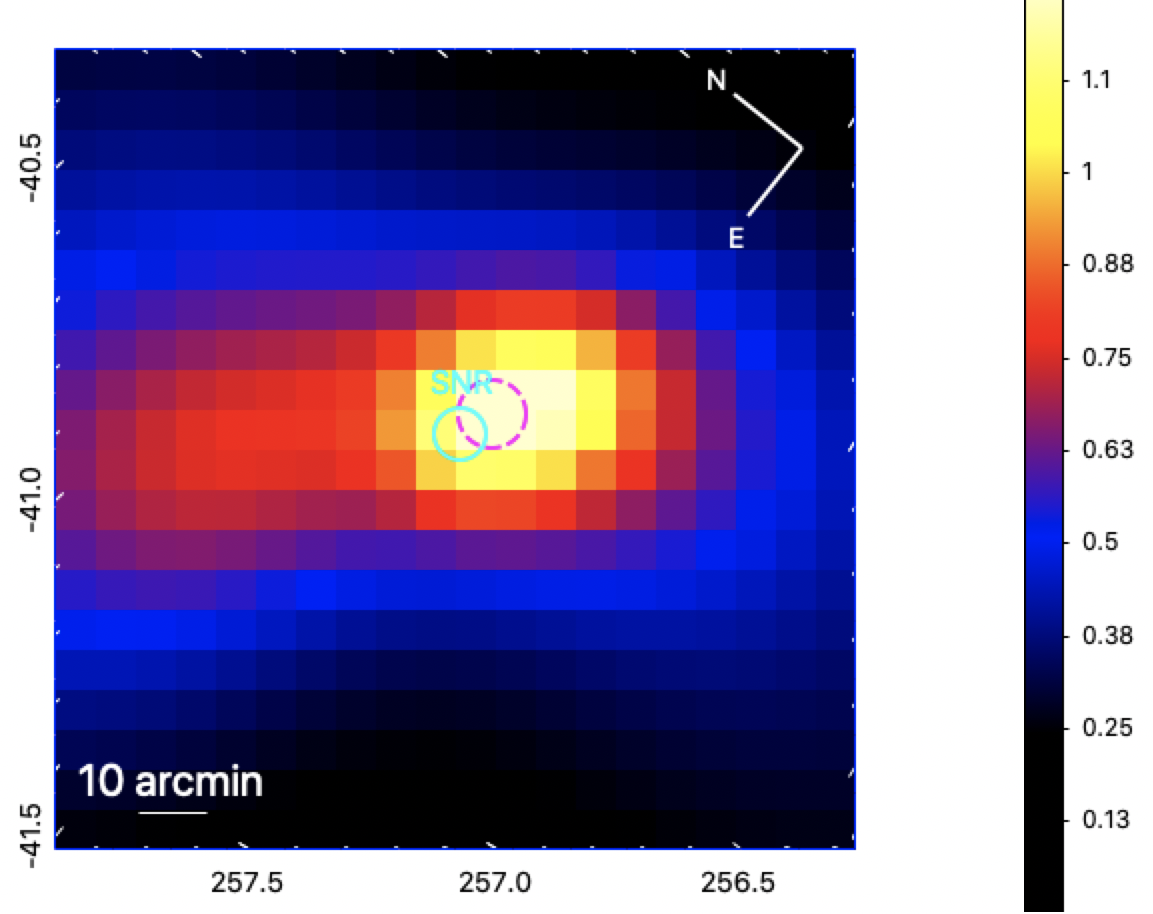}
\end{minipage}
\caption{{ {\it Left:} $\gamma$-ray SED of 2FHL~J1703.4--4145, using data from the 3FHL catalog \citep{ajello2017}. Also plotted is the data from the likely TeV counterpart HESS~J1702-420 \citep[see text,][]{hess2018}.
\textit{Right}: An adaptively smoothed $\gamma$-ray image of the \fermi\ $\geq$50\,GeV sky where 2FHL~J1703.4$-$4145 is located. The source shows no significant evidence of extended emission. The count map is not background subtracted and the 2FHL source is embedded in Galactic diffuse emission as it is located along the Galactic plane. This is the same map from \citet[][see Figures 1 and 5]{ackermann2015}. The angular diameter in radio of the SNR is indicated in cyan and the 95\% uncertainty position of the 2FHL source is marked by the dashed magenta circle.}
}\label{fig:sed_gamma_map}
\end{figure*}

2FHL~J1703.4--4145 was first detected at $>$50\,GeV in the 2FHL catalog and presents a particularly hard $\gamma$-ray spectrum with photon index $\Gamma_\gamma=1.24\pm0.36$ and a maximum photon energy  of $\sim$1.7\,TeV \citep{ackermann2015}. The source was also detected above 10\,GeV and reported in the 3FHL catalog\footnote{We will continue to use the 2FHL identifier for 2FHL~J1703.4--4145 for two reasons. One being that the 2FHL and 3FHL identifiers and properties for 2FHL~J1703.4--4145 are the same and the other being that the sub-sample this source is a part of was generated using the unidentified objects in the 2FHL catalog as described in Section 1. Any major differences between 2FHL sources in this Galactic sample and subsequent \fermi\ catalog counterparts will be addressed in their respective reports.} (see Figure \ref{fig:sed_gamma_map}). The 3FHL catalog \citep{ajello2017} covers a larger energy range from 10\,GeV to 2\,TeV  and adopts a longer exposure of 84 months, compared to the 80 months used in the 2FHL catalog. Furthermore, the 3FHL catalog results from taking full advantage of improvements provided by Pass 8, using the point-spread-function (PSF)-type event classification, improving sensitivity and leading to an increase in photon counts 10 times greater than what is reported in the 2FHL catalog. Therefore, the 3FHL counterpart provides us with more information of 2FHL~J1703.4-4145 in the high energy regime. Above $\gtrsim$10\,GeV, the \fermi\ point spread function (PSF) is $\sim0.1^{\circ}$ at 68\%\,confidence level \citep[C.L.,][]{ajello2017}. The source is compact and shows no clear evidence of extended emission beyond the PSF of the \textit{Fermi}-LAT in this energy range. 

To further investigate the properties of this VHE object, we performed an X-ray analysis on \xmm\ archival data from 2001 (ObsID: 0111210401, PI: M. Watson) in order to better understand which part of the SNR is likely responsible for the $\gamma$-ray emission observed.

\section{\xmm\ X-ray analysis}\label{sec:xdata}
\subsection{\xmm\ Data Reduction and Analysis}
\begin{figure}%[t!]
\centering
\includegraphics[width=0.4\textwidth]{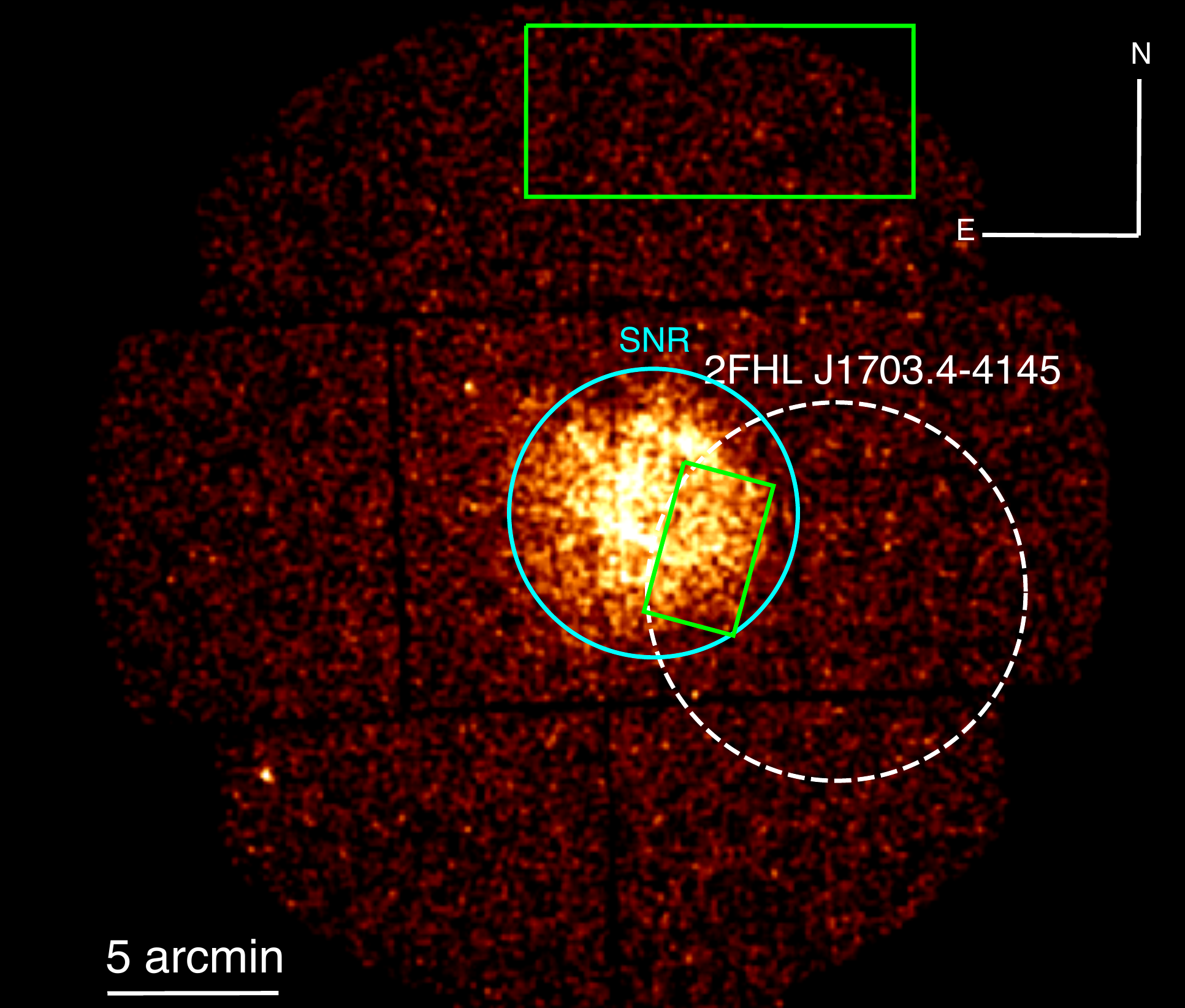}
\caption{Smoothed MOS2 0.5--10\,keV image of SNR~G344.7--0.1 which is indicated in cyan. The white dashed circle ($r$=5.2$^\prime$)  represents the 95\% positional uncertainty of 2FHL~J1703.4--4145}. The green box covering the Western half of the SNR is used for the source spectrum and the large green rectangle at the top is used for the background spectrum.\label{fig:xray}
\end{figure}

The entire system of SNR~G344.7--0.1 was observed with \xmm\ in 2001 for 12.2\,ks in full frame mode. In Figure \ref{fig:xray} we show the smoothed 0.5--10\,keV image of SNR~G344.7--0.1, as seen with the MOS2 camera mounted on \xmm. Data reduction was performed using SAS software (v17.0) with the corresponding calibration files for \xmm\footnote{\xmm\ calibration files can be found here \url{https://heasarc.gsfc.nasa.gov/docs/xmm/xmmhp_caldb.html}.}. In Figure \ref{fig:xray}, the position of the SNR is indicated using the 8$^\prime$ radio angular diameter as well as the location of 2FHL~J1703.4-4145 with respect to the SNR. The 2FHL 95\% confidence region overlaps the western half of the supernova remnant which suggests that the $\gamma$-ray emission is associated to this object.
%{\color{red}{\texttt {add here more information about data reduction.}}}

The selected regions for the spectral fitting process are indicated in Figure \ref{fig:xray}. Modeling both the source and background, we perform a spectral fitting on the resulting spectra from MOS1, MOS2, and PN using the HEASOFT software package \citep[v6.19,][]{drake2016} with XSPEC (v12.9.1) in order to find the best model to characterize the observed emission. The background is modeled taking into account both the instrumental and astrophysical contributions. The first is modeled as a combination of quiescent soft protons, CR-induced continuum, and fluorescence lines and the latter is modeled as both the emission from the Galactic halo and the cosmic X-ray background\footnote{For a careful treatment of faint, extended objects using the same background model see Paper Ia and references therein.}. The instrumental background component is negligible across the energy range but does account for prominent fluorescence lines including those of Al-K ($\sim$1.5\,keV), Si-K ($\sim1.8$\,keV), Cr-K ($5.4-5.9$\,keV), Mn-K ($\sim5.9$\,keV), and Fe-K ($\sim6.4$\,keV). We binned the spectrum at 25 counts per bin and used the C-stat statistic.

%The spectral fitting of bright, point-like X-ray sources can be performed subtracting the background emission since the signal-to-noise ratio of the source is large enough that removing a small fraction of counts from the fitted spectrum does not affect the quality of the analysis. If the object is exceptionally faint or extended, it is more beneficial to model the background instead \citep[see e.g.,][]{leccardi, eagle2019}. The background modeling approach is demonstrated in Paper Ia. Here we adopt the background subtraction method.  

\subsection{Spectral Analysis Results}
\begin{figure}[!t]
\centering
\includegraphics[width=1.05\linewidth]{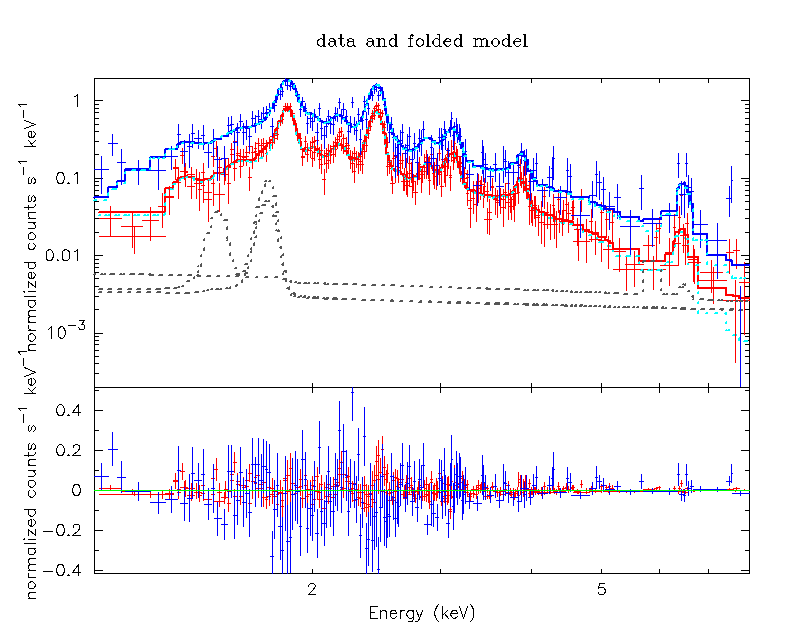}
\caption{\textit{Top}: XMM-\textit{Newton} MOS1 and MOS2 (red), and pn (blue) data of the source and background regions (see Figure \ref{fig:xray}) and the best-fit model (blue and red solid lines) obtained using \texttt{phabs(vnei + gaussian)}. The emission lines observed between 1\,keV and 7\,keV correspond to metals in the plasma: Mg ($\sim$1.5\,keV), Si ($\sim$1.7\,keV), S ($\sim$2.4\,keV), Ar ($\sim$2.5\,keV), Ca ($\sim$4\,keV), and Fe ($\sim$6.4\,keV) which is consistent with the results reported in \citet{giacani2011, yama2011} and \citet{yama2020}. The dotted dark grey lines correspond to the instrumental background contributions for each data set. The dashed cyan lines correspond to the combination of source and astrophysical background contributions. \textit{Bottom}: Residual map of the data and best fit model.
}\label{fig:vnei}
\end{figure}

SNR~G344.7-0.1 has been studied in the X-rays in \citet{combi2010, giacani2011}, \citet{yama2011}, and \citet{yama2020}, all finding that thermal models can describe the entire remnant appropriately. Using this as a starting point for our spectral analysis, we found that the \texttt{vnei} model \citep{chev1983} best describes the spectrum from the western half of the SNR. The \texttt{vnei} model characterizes a non-equilibrum, ionized and collisional plasma that is constant in temperature and allows the ionization timescale to vary. The best fit parameters for the model are reported in Table \ref{tab:vnei} and the best fit model is shown in Figure \ref{fig:vnei}. There are several metal emission lines present including: Mg ($\sim$1.5\,keV), Si ($\sim$1.7\,keV), S ($\sim$2.4\,keV), Ar ($\sim$2.5\,keV), Ca ($\sim$4\,keV), and Fe ($\sim$6.4\,keV). These findings are consistent with previous works \citep{yama2011, yama2020} and the Fe emission observed in the spectrum supports the scenario where the SNR may result from a type Ia SN \citet{yama2011}.

The emission lines are modeled using \texttt{vnei} where available but we find a Gaussian component can better model one of these emission lines and provide overall better statistics. The feature to require additional modeling is the Fe-K$\alpha$ line at 6.4\,keV as this is not well modeled with \texttt{vnei}, likely due to this arising from another plasma component with a different ionization timescale \citep{yama2011, yama2020}. %Moreover, the effective area of the \xmm\ instrument declines around 10\,keV and combined with the very low data quality apparent beyond 8\,keV, this makes the Fe line difficult to model. %Therefore, an upper limit on the Fe line emission is found to be the best-fit value (see Table \ref{tab:vnei}). 
%The second Gaussian component is for a very weak Mg emission line at 1.47\,keV, which is not modeled well in \texttt{vnei}, similar to Fe, likely also arising from another plasma component. In fact, \citet{yama2011} reports a similar finding for Mg presence and suggests it results from a combination of both shocked ejecta and swept-up ISM. Modeling both with Gaussian components,

We find the best fit for the observed X-ray spectrum to be an absorption component \texttt{phabs} \citep{phabs} multiplying \texttt{vnei} and a \texttt{gaussian}. The solar abundances are assumed to be the \citet{wilms2000} ones. The \texttt{vnei} parameters H, He, C, N, and O are set to unity, following \citet{yama2011}. Our results are consistent with \citet{giacani2011} using the same observation from \xmm\ and the same model. However, comparing the X-ray data from \xmm\ to {\it Suzaku} and {\it Chandra} X-ray data \citep{yama2011, yama2020} shows that the Fe line cannot be attributed to background fluorescence alone as a Fe K line is clearly detected in all data sets. We have carefully modeled the background, including the Fe K$\alpha$ fluorescence at 6.4\,keV. Significant Fe emission from the SNR is apparent and is therefore included in the source model. Contributions from both background and source emission are plotted separately in Figure \ref{fig:vnei}. $N_H$ values reported here are slightly higher than previous works \citep{giacani2011,yama2011} and this is due to assuming \citet{wilms2000} abundances as in \citet{yama2020}. Lastly, we note the super-solar abundances of the metals, with the exception of Mg, indicating this emission is from a mixture of shocked SN ejecta and swept-up ISM, similarly found in \citet{giacani2011, yama2011, yama2020}. 

%In conclusion, the complexity of the spectral fitting suggests we are seeing emission from both shocked SN ejecta and swept-up ISM and, combined with limited photon statistics, cannot easily model all components uniformly.

%\begingroup
%\renewcommand*{\arraystretch}{1.2}
%\begin{table*}
%\centering
%\scalebox{1.}{
%\begin{tabular}{|c c c |}
%\hline
%\ $\chi^2$ & d.o.f.$^a$ & Reduced $\chi^2$ \
%\\
%\hline
%944.92 & 800 & 1.18 \\
%\hline
%\end{tabular}}\caption{Summary of the best-fit statistics. {\footnotesize %\textit{$^a$ degrees of freedom}}}
%\label{tab:vnei2}
%\end{table*}
%\endgroup
%\todo[inline]{Update parameters and uncertainties.}
\begingroup
\renewcommand*{\arraystretch}{1.2}
\begin{table*}
\centering
\scalebox{1.}{
\begin{tabular}{|c c c |}
\hline
\ C--stat & d.o.f.$^a$ & C--stat/d.o.f. \
\\
\hline
850.22 & 719 & 1.18 \\
\hline
\hline
\ Component & Parameter & Best-Fit Value \
\\
\hline
\ {\texttt{phabs}}$^b$ & N$_{H}$(10$^{22}$ cm$^{-2}$) & 6.81$_{-1.20}^{+1.03}$ \
\\
\hline
\ \texttt{vnei} & $kT$(keV) & 1.33$_{-0.22}^{+0.34}$ \\
\ & Mg & $<0.83$ \\
\ & Si & 2.70$_{-0.91}^{+1.01}$ \\
\ & S & 3.10$_{-0.44}^{+0.76}$ \\
\ & Ar & 3.00$_{-0.58}^{+0.63}$ \\
\ & Ca & 5.40$_{-1.35}^{+1.66}$ \\
\ & $\tau$ ($10^{11}\,$cm$^{-3}$s) & 1.05$_{-0.26}^{+0.35}$ \\
\hline
\ Fe & E(keV) & 6.47$_{-0.045}^{+0.048}$ \\
\ & $\sigma$ (keV) & 0.12$_{-0.050}^{+0.070}$ \\
\ & Normalization & 3.86$_{-1.03}^{+1.23}\times10^{-5}$\\
\hline
\end{tabular}}
\caption{Summary of the statistics and parameters for the best-fit model in our analysis, \texttt{phabs(vnei + gaussian)}. Metal abundances are reported in solar units. \footnotesize{$^a$ Degrees of freedom, $^b$ Absorption cross section set to \citet{verner1996}.}}
\label{tab:vnei}
\end{table*}
\endgroup

\section{Multi-wavelength Information}\label{sec:multi}
\subsection{Radio}

\begin{figure}[!t]
\centering
\includegraphics[width=0.90\linewidth]{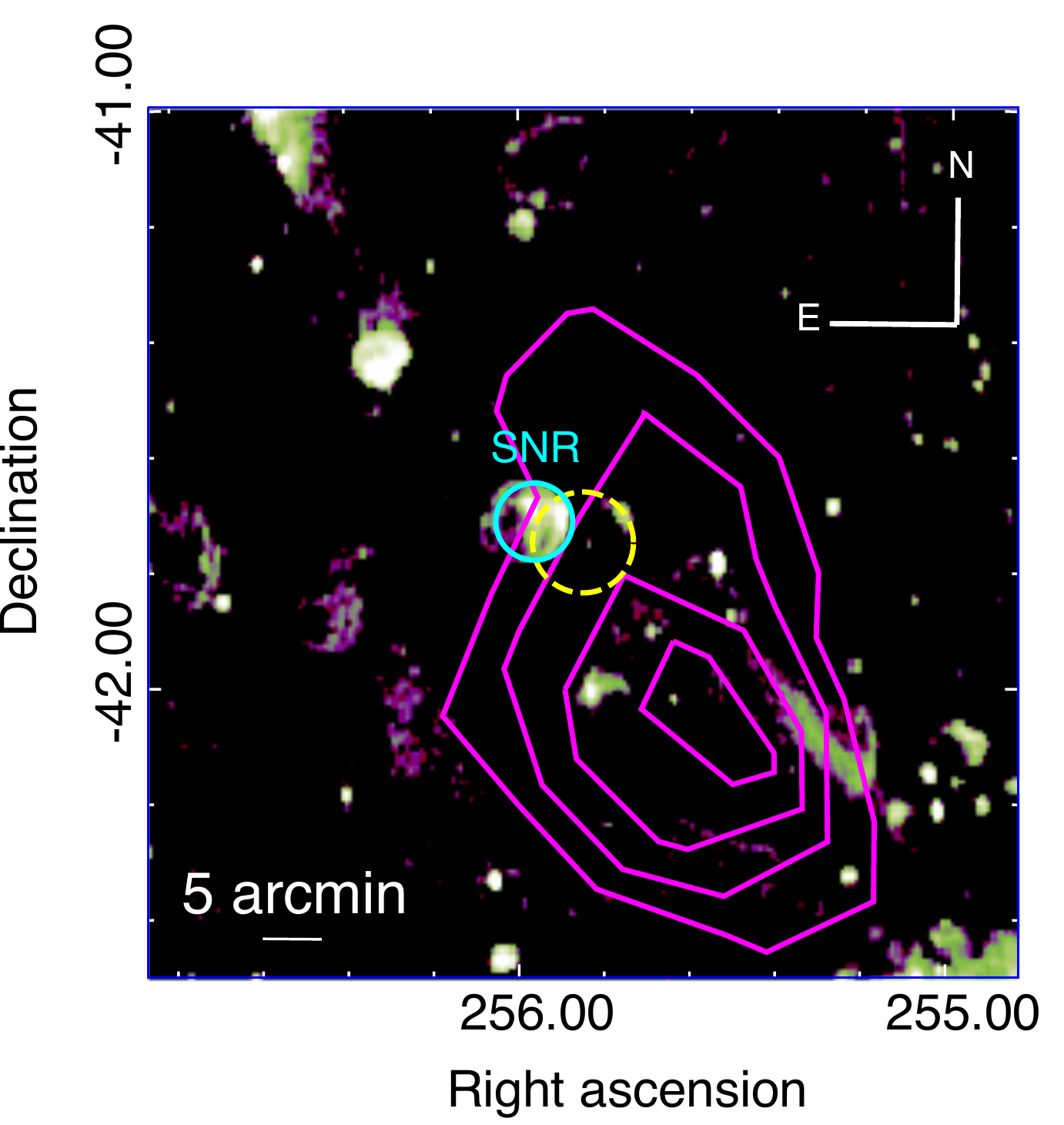}
\caption{Molonglo Observatory Synthesis Telescope (MOST) 843\,MHz map from \citet{whiteoak1996} where the SNR shell of 8$^\prime$ is readily seen and marked by the solid cyan circle with the 95\% confidence region of 2FHL~J1703.4--4145 indicated by the yellow dashed circle. We plot HESS~J1702--420 flux contours from \citet{hess2018} to show where the emission peak is located.} We note that the bright anomaly just within the Northwest corner of the \fermi\ uncertainty region is more than likely a young stellar object and not associated to the observed $\gamma$-ray emission (see text).\label{fig:radio}
\end{figure}

Central emission within the shell is clearly visible in many radio bands including 408 and 5000\,MHz \citep{caswell1975}, 1465\,MHz \citep{dubner1993, giacani2011}, 115\,GHz \citep[$^{12}$CO][]{giacani2011}, 1.4\,GHz \citep[ATCA and VLA][]{giacani2011}, and 843\,MHz \citep[see Figure \ref{fig:radio},][]{whiteoak1996}.

\citet{dubner1993} described the SNR in 1465\,MHz as having a revealing shell morphology and reported for the first time the detection of bright central emission indicating the remnant to be a composite SNR type. \citet{dubner1993} also provided the first linear diameter and distance estimations of $D\approx33$\,pc and $d\approx14$\,kpc based on the $\Sigma-D$ calibration technique from \citet{huang1985}. This technique is not very reliable due to large intrinsic dispersion \citep[see e.g.][and references therein]{dubner1993, yama2011}, but is considered a conservative distance estimate for the SNR to date \citep{yama2011}. 

As can be seen in Figure \ref{fig:radio}, the western and central part of the SNR are much brighter than the eastern half. 1.4\,GHz data from the Australia Telescope Compact Array (ATCA) and the Very Large Array (VLA) reveal the shell to be nearly complete with a diameter of 8$^\prime$ \citep{giacani2011}. The observed central X-ray emission (Figure \ref{fig:xray}) is totally encompassed by the radio shell. Furthermore, the radio surface brightness is also seemingly correlated to the IR emission that peaks to the West as well at $24$\,$\mu$m (e.g., see Figure \ref{fig:composite}).%\footnote{The correlation between the varying wavelengths with the SNR morphology is discussed in more detail in section 3.1 of \citet{giacani2011}.}. 

A shock-cloud interaction is a preferred scenario to explain the bright central radio emission as well as other enhanced emission \citep{combi2010, giacani2011}. This is supported by the broad presence of both neutral hydrogen and carbon monoxide\footnote{Public data for the Mopra CO survey data release 3 can be found here: \url{doi:10.7910/DVN/LH3BDN} and was first reported in  \citet{mopra2018}.}  \citep{giacani2011,lau2019, yama2020}. 

We note that the bright anomaly just within the northwest corner of the 95\% \fermi\ uncertainty region (see Figures \ref{fig:radio} and \ref{fig:composite}) is more than likely a young stellar object (YSO) and not associated to the observed $\gamma$-ray emission\footnote{This anomaly is bright in both the radio and IR and is coincident with an HII region (HRDS~G344.593--00.044), a sub-millimeter YSO (AGAL~G344.606--0.031), and a large HII bubble (SPK2012~MWP1G344590--00500) and therefore is more than likely unrelated to the SNR and consequently the 2FHL source.}.
{\color{black}

\subsection{Infrared}
{\it Spitzer} GLIMPSE survey data of SNR~G344.7--0.1 at $24$\,$\mu$m is shown in Figure \ref{fig:composite}, left panel. 
%\citet{spitzer2011} find that SNR~G344.7--0.1 has the highest temperature of shocked dust of their MIR study sample with a temperature of 66\,K. Additionally, 
It is reported in \citet{spitzer2011} that the SNR exhibits several features in the IR band that are indicative of an interaction between the SNR shock wave and its dense surroundings. %that molecular hydrogen is detected in the MIR spectrum which suggests that the main cooling mechanism for the SNR is in the form of IR continuum cooling from the dust. Molecular hydrogen emission is also an indicator for an interaction between the forward shock and a molecular cloud \citep{spitzer2011}.
%Based on the Ne III and Ne II ratio of 0.54, \citet{spitzer2011} also conclude the forward shock velocity to be $\sim280$\,km s$^{-1}$ and an electron density $n_e = 100$\,cm$^{-3}$.

The IR image at 24$\mu$m provides indicators of shocked dust being swept up by the forward shock of the remnant as it expands into the ISM. The IR filament observed to the West of the remnant shell and well within the 2FHL uncertainty region could indicate where a shock-cloud interaction is occurring that may be accelerating particles to cosmic ray energies. To support this claim, there is an abundance of gas, particularly HI, in the region of the SNR that could provide a dense medium for the forward shock to run into \citep[see previous Section and][]{combi2010, giacani2011, yama2011, lau2019}. Moreover, it is discovered a bright X-ray filament in the 1.76--1.94\,keV energy range as seen with {\it Chandra} coincides with the IR filament \citep{yama2020}. However, it is suggested to be of SN ejecta origin rather than a forward shock front propagating into the ISM due to the relative abundances of S, Ar, Ca, and Si being comparable to solar values accompanied by a much lower presence of Mg. On the other hand, other recent work describes the dust features being consistent with a non-SN origin \citep[i.e., swept-up material, see][]{chawner2019, chawner2020}.

Moreover, there is enhanced mid-IR emission from shocked ionized gas that coincides with the bright radio central emission \citep{chawner2019}. IR emission is also detected in the north as well, pointing to an interaction between the supernova shock and a molecular cloud in front of the SNR \citep{chawner2019}. 

\begin{figure*}[htbp]
\begin{minipage}[b]{.5\textwidth}
\hspace{3cm}
\centering
\includegraphics[width=0.6\linewidth]{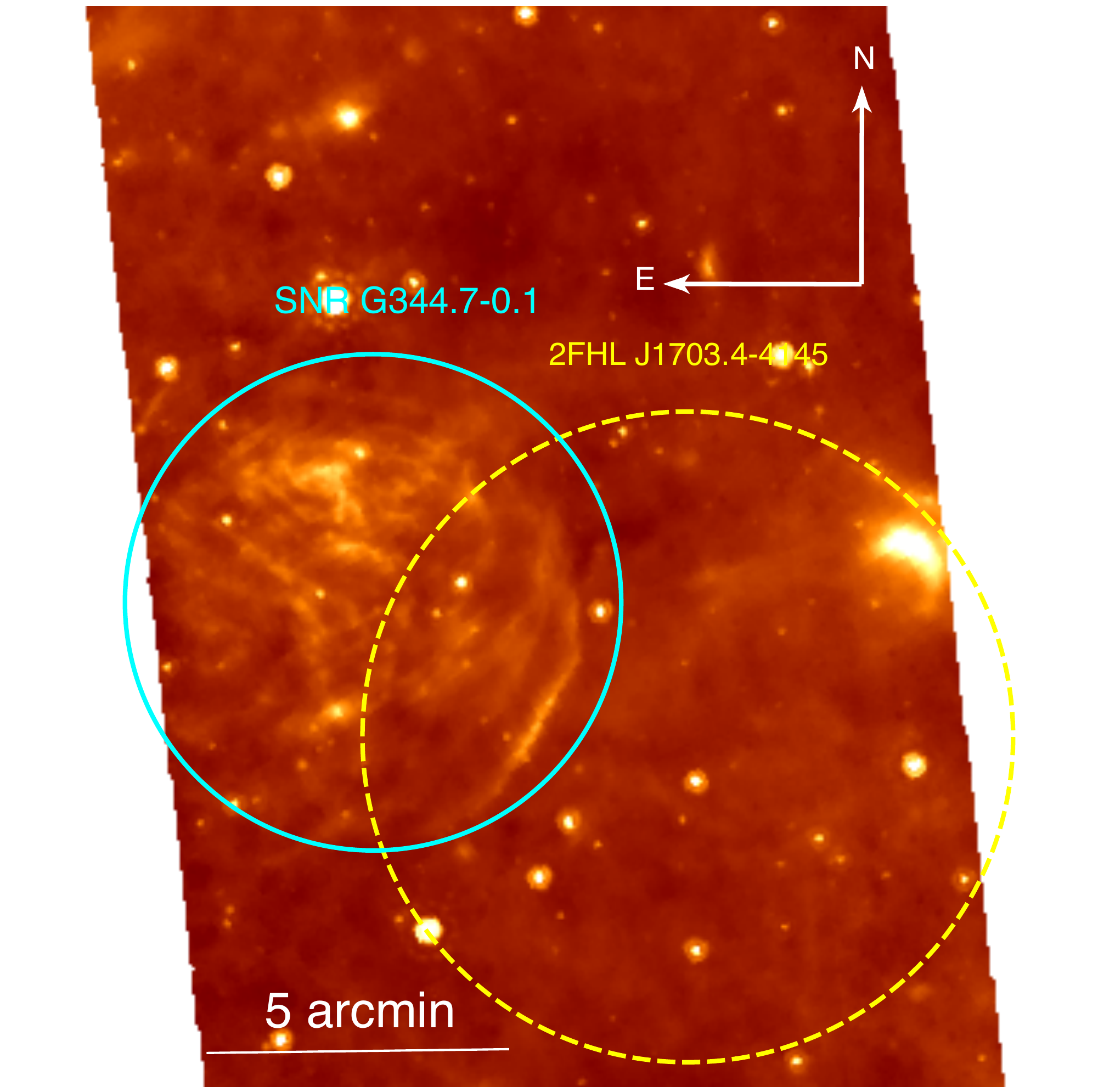}
\end{minipage}
\begin{minipage}[b]{.5\textwidth}
\hspace{-.4cm}
\centering
\includegraphics[width=0.68\linewidth]{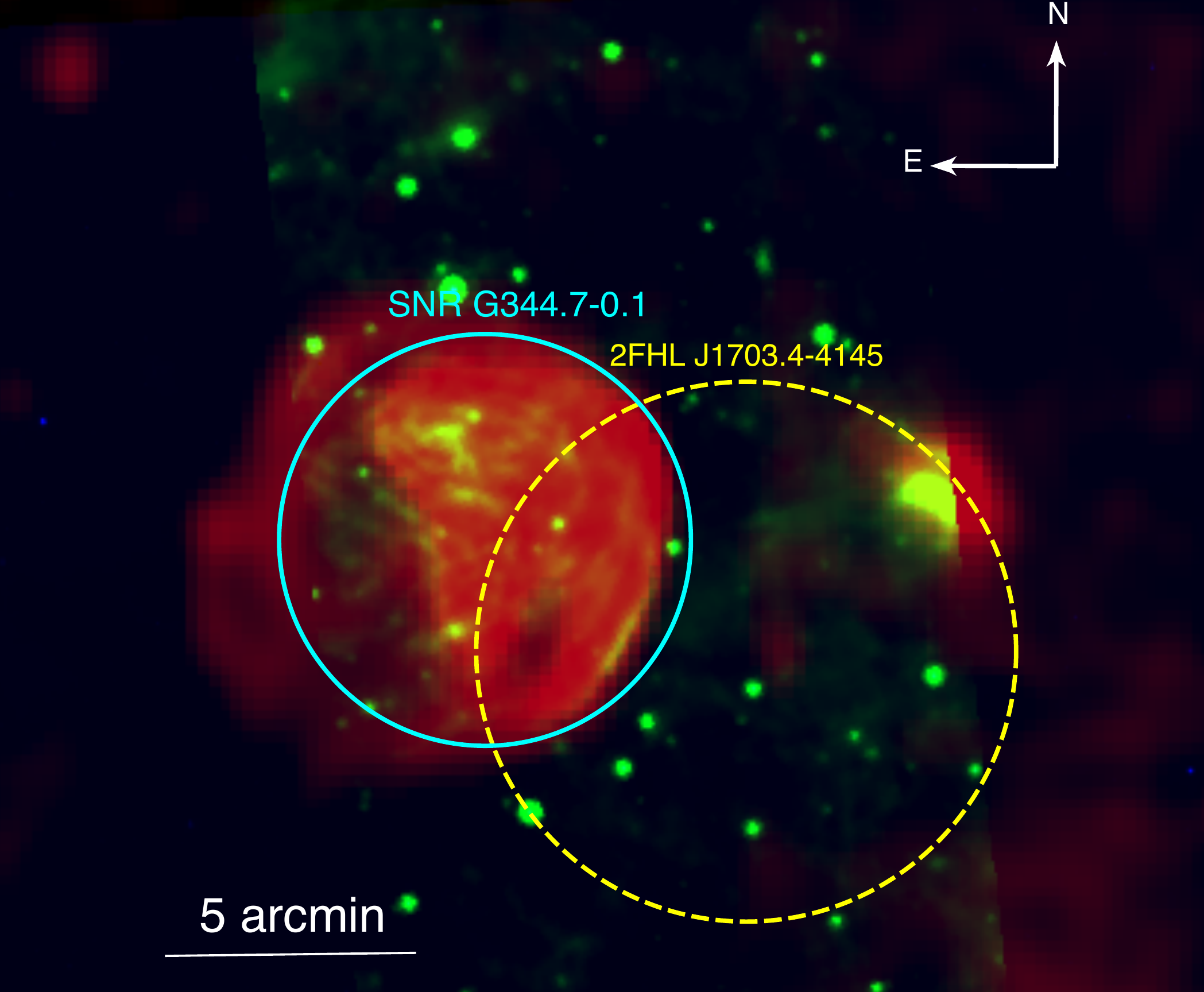}
\end{minipage}
\caption{{\textit{Left}}: SPITZER 24$\mu$m image of SNR~G344.7--0.1 with 2FHL~J1703.4--4145 95\% uncertainty region indicated. Note the bright filament on the western edge of the SNR that overlaps well into the 2FHL region.
{\textit{Right}}: 843\,MHz (red) superposed with 24$\mu$m (green) and 8$\mu$m (blue) SPITZER MIPS and IRAC images, respectively. The bright radio emission is paired with bright IR emission in the center of the SNR. A bright IR filament traces the western edge of the SNR, coincident with the position of \fhl. These two points indicate an interaction between the remnant and its surroundings. }\label{fig:composite}
\end{figure*}

\subsection{Soft X-rays}
In the soft X-rays, SNR~G344.7--0.1 is encompassed with thermal emission across the remnant within the radio shell as seen with \xmm, {\it Chandra}, {\it Suzaku}, and ASCA \citep[see previous Section, Figure \ref{fig:xray}, and][]{yama2005,combi2010,giacani2011,yama2011, yama2020}. The temperature of the SNR is roughly $kT\sim$\,1.0--1.5\,keV across the remnant \citep[see Section \ref{sec:xdata} and][]{combi2010,giacani2011, yama2011, yama2020}. Specifically in this work, the X-ray emission overlapping with the observed $\gamma$-ray emission is found to have a temperature $kT=1.33_{-0.22}^{+0.34}$,keV at 90\%\,C.L.

%\subsection{Gamma-rays}
%Extended and yet unidentified source, HESS~J1702--420, located at (R.A., Dec.)$=(255.63\degree, -42.07\degree) \pm0.05\degree$ in J2000, is a likely TeV counterpart of 2FHL~J1703.4--4145 due to their positional coincidence with SNR~G344.7--0.1 and compatible $\gamma$-ray spectral energy distributions \citep[see Figure \ref{fig:sed_gamma_map} and][]{hess2018}.

%Several possibilities of the connection between SNR~G344.7--0.1 and HESS~J1702--420 have been explored. In summary, the role HESS~J1702--420 plays in association with the SNR and 2FHL~J1703.4--4145 is complex but a favored scenario is one where the SNR forward shock is interacting with a cloud to the west, accelerating particles to CR energies that then escape and diffuse into the ISM. The CRs bombard an as of yet unidentified, large molecular cloud that is coincident with the position of the HESS~J1702--420 emission peak. 
%is not well understood and could still be misinterpreted and could instead be from a separate source altogether. 
%In section \ref{sec:discuss}, we discuss further implications from broadband SED modeling which seem to point toward the described scenario.

%{\color{red} Could the SED modeling support the TeV halo scenario? If ICS? Yes but we see no X-rays beyond SNR.}.

\section{Discussion}\label{sec:discuss}}

\subsection{Efficient Particle Acceleration}
SNRs are widely thought to accelerate a significant fraction of Galactic CRs through diffusive shock acceleration (DSA) in their high-velocity blast waves. The $\gamma$-ray emission in the MeV-GeV band from regions with relatively high ambient density is expected to be hadronic in origin \citep[see][Figure 6, for an example]{castro_2013b}, and hence evidence of CR hadron acceleration at these shocks. Proton-proton collisions between shock accelerated CR ions and ambient protons are enhanced in high-density regions such as an interaction between a SNR forward shock and a molecular cloud or a CR accelerator located near a high density cloud. The energetics of 2FHL~J1703.4--4145 make this site another promising source for efficient particle acceleration to CR energies. The best-fit physical parameters for both leptonic and hadronic emission scenarios are investigated in \S \ref{sec:modeling}.

The X-ray band may also provide clues. Where the shock has become radiative, it is likely to become bright in the GeV band, such as W44 and IC 443, and the resulting X-ray emission is commonly characterized by a center-filled X-ray morphology, rather than a shell-like one, similar to what is observed for SNR~G344.7--0.1 \citep{castro2015}. Generally bright optical filaments that are associated with thermal X-ray emission provide evidence for the shock to still have enough speed to heat the surrounding medium to X-ray emitting temperatures and hence a significant part of the shock could likely be non-radiative. Thus if bright optical filaments can be discovered in the region of 2FHL~J1703.4--4145 then it is possible CRs in this region have been produced through DSA.

\subsection{Modeling Spectral Energy Distribution}\label{sec:modeling}
The multi-wavelength information available can be combined to build a picture of the broadband spectral characteristics of the region. Assuming the GeV $\gamma$-ray emission of \fhl\ is indeed the result of radiation from a relativistic particle population accelerated by the SNR~G344.7--0.1 shock, it is possible to model the broadband emission from the shock-accelerated non-thermally distributed electrons and protons and hence derive constraints on the physical parameters of the mechanism responsible for the observed emission. %The data of the region are shown in figure \ref{fig:spec}, where we plot the 3FHL data from \citet{ajello2017} and the TeV $\gamma$-ray data is from the H.E.S.S. Galactic Plane Survey \citep{hess2018}.

We assume the distribution of the accelerated particles in momentum to be %$dN_{i}/{dp} = a_{i} \,p^{-\alpha_{i}} \exp\left(-{p}/{p_{0\,i}}\right)$.
\begin{equation}
\frac{dN_{i}}{dp} = a_{i} \,p^{-\alpha_{i}} \exp\left(-\frac{p}{p_{0\,i}}\right)
\label{eq:uno}
\end{equation}
Here, subindex $i$ represents the particle type (proton or electron), and $\alpha_{i}$ and $p_{0\,i}$ are the spectral index and the exponential cutoff momentum of the distributions. The coefficients for the particle distributions, $a_{p}$ and $a_{e}$, are set using the total energy in relativistic particles and the electron to proton ratio as input parameters, together with the spectral shape of the distributions. The spectral indices of electron and proton distributions are assumed to be equal and for the non-thermal radiation from these particle distributions we have used $\pi^0$-decay emission from \citet{kamae_2006,mori_2009} and inverse Compton (IC) emission from \citet[][and references therein]{baring_1999}. For more details on the model and application see \citet{castro_2013} and Paper Ia.

\begin{figure*}[htbp]
\begin{minipage}[b]{.5\textwidth}
\hspace{3cm}
\centering
\includegraphics[width=1.1\linewidth]{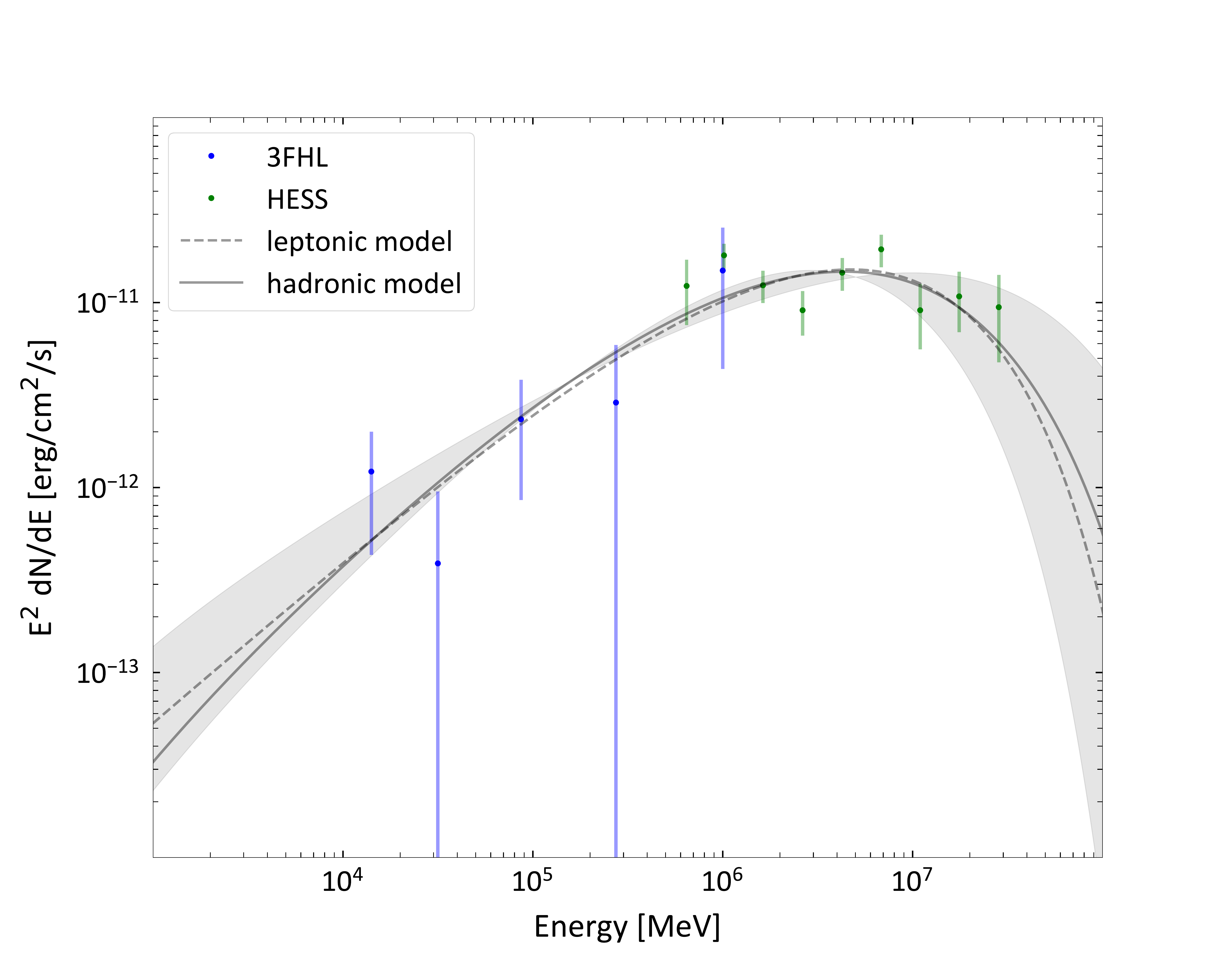}
\end{minipage}
\begin{minipage}[b]{.5\textwidth}
\hspace{-.4cm}
\centering
\includegraphics[width=1.1\linewidth]{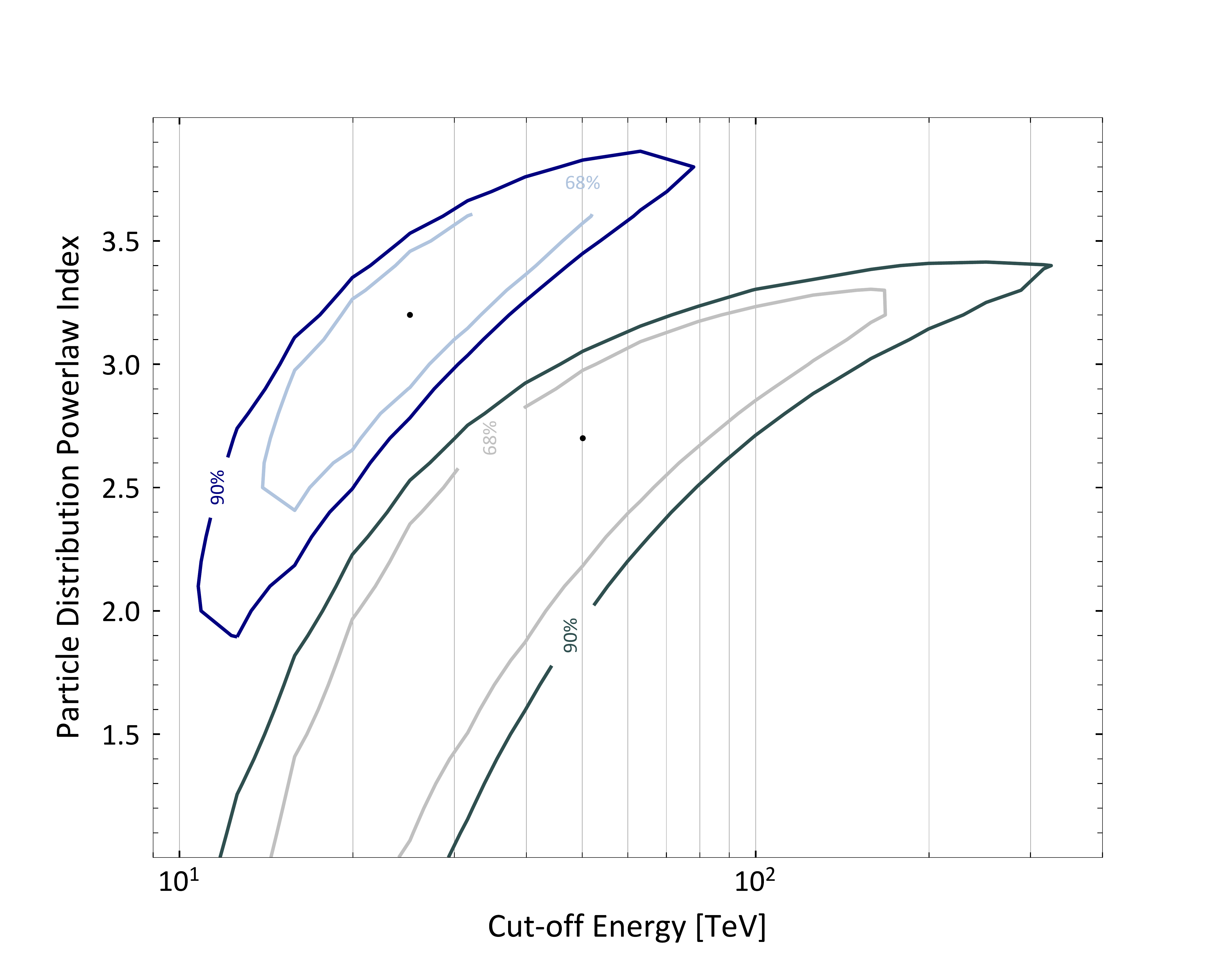}
\end{minipage}
\caption{{\textit{Left}}: The spectral energy distribution model constrained to 3FHL and HESS spectral data. The solid grey line (hadronic scenario) and the dashed grey line (leptonic scenario) demonstrate the resultant $\gamma$-ray spectrum of radiation from relativistic protons or electrons, respectively.
{\textit{Right}}: IC decay (blue) and pion decay (grey) model contour plot for the spectral fitting results, marking the 1$\sigma$ and 2$\sigma$ uncertainties. The black dot shows the best-fit values. The best-fit values are listed in Table \ref{tab:models}.}\label{fig:contours}
\end{figure*}

We use the model outlined above to establish the approximate ranges of some of the physical parameters that would result in emission that fits the \fermi\ data, as well as complying with available data at other wavelengths. %The common parameters for all models considered are a relativistic electron to proton ratio of $k_{ep}=0.01$ (determined in observations of CR abundances on Earth), and a shock compression ratio of 4. Both of these standard assumptions are discussed in more depth in \citet{castro_2013}. Additionally, 
We adopt a distance of $d=6.3$\,kpc. %, and fix the spectral index of the relativistic proton and electron distributions in momentum to be $\alpha_i=4$. This last assumption is adopted given that neither the radio nor the $\gamma$-ray observations allow for a tightly constrained spectral index of the emission spectrum and hence, we default to the canonical spectral index expected from diffusive shock acceleration \citep[see ][and references therein]{reynolds_2008}. 
The input parameters for each model considered are included in Table \ref{tab:models} and the resulting SED models are shown in Figure \ref{fig:contours}, left panel. 
%{\color{red} Discussion of models blah blah blah....}

%since analytic and semi-analytic models of particle acceleration at shocks suggest this is the case \citep{reynolds_2008}. For the non-thermal radiation from these particle distributions we have used $\pi^0$-decay emission from \citet{kamae_2006,mori_2009}, synchrotron and inverse Compton (IC) emission from \citet[][and references therein]{baring_1999}, and non-thermal bremsstrahlung emission from \citet{bykov_2000}. For more details on the model for the particle distribution and their simulated emission see \citet{castro_2013}. 

\begin{table}
\centering
\begin{tabular}{ccccccccc}
\hline
\hline
\noalign{\vskip 1mm} 
%\rule{0pt}{2.6ex}

\multirow{2}{*}{}&\multirow{2}{*}{}&$p_{\text{0}}$&$\Gamma_\gamma$&$E_{\text{CR}}$ \\

&&{(TeV/c)}& & {($10^{50}$ erg (d/6.3\,kpc)$^2$)}\\
%\rule{0pt}{2.6ex}
\noalign{\vskip 1mm} 
\hline
\noalign{\vskip 1mm} 
%\rule{0pt}{2.6ex}
\multirow{2}{*}{\it Leptonic}  & &  25.1   &  2.2  & 0.0093 \\ 
%\rule{0pt}{2.6ex}
\noalign{\vskip 3mm} 
\hline
\noalign{\vskip 2mm} 
\multirow{3}{*}{\it Hadronic}  & &  50.1  & 1.6 & 4.3 \\
\\

% \multirow{2}{*}{\it Leptonic}&{\it A}  &  5.3   & 2.8  & --   & 8.8 \\ 
% &{\it B}  &  28    & 9.2  & --   & 1.1 \\ 
% %\rule{0pt}{2.6ex}
% \noalign{\vskip 1mm} 
% \hline
% \noalign{\vskip 1mm} 

% \multirow{3}{*}{\it Hadronic}&{\it C}  &  10    & 48   & 5.93 & -- \\ 
% &{\it D}  &  1.5   & 48   & 5.93 & -- \\ 
% &{\it E}  &  600   & 12   & 1.5  & -- \\ 

\noalign{\vskip 1mm} 
\hline
\noalign{\vskip 1mm} 

\end{tabular}
\caption{{\small{\uppercase{SED Input Model Parameters. --}}}: Target density in hadronic model is assumed to be 1\,cm$^{-3}$.}
\label{tab:models}
\end{table}

%{\large\color{red}Where Jordan has left off!}

%Since the multi-wavelength information only provides upper-limits, we use the model described above to roughly estimate the ranges that some parameters can take while still fitting the \fermi\ data.  

When trying to fit the broadband spectral data, including radio at 1.4\,GHz from ATCA \citep{giacani2011}, X-ray (from {\it Chandra} data, ObsID: 21117, PI: Yamaguchi), and $\gamma$-ray \citep[from][]{ajello2017, hess2018}, it becomes apparent that the radio and X-ray data are not connected to the $\gamma$-ray emission mechanism, suggesting that two different electron particle populations are at work.  Fitting the multi-wavelength data using one electron population requires the magnetic field to be much lower than the average ISM value of $\sim3\,\mu$G which is unlikely. Additionally, including the $\gamma$-ray data indicates that it is unlikely the CRs responsible for the GeV-TeV emission are trapped in the SNR shock. If this were the case, then considering the typical minimum magnetic field strength in the Galaxy being approximately $\sim3\,\mu$G and that it would be compressed by the shock at least by a factor of 4 (compression factor for a normal SNR shock), we would expect to detect a flux of non-thermal X-ray emission larger than allowed by the upper limit we have derived from the \xmm\ data. The only scenario that fits the radio, X-ray and $\gamma$-ray data is one where the $\gamma$-ray emission is a result of CRs at the SNR shock where the local density is in the order of $\sim$1,000\,cm$^{-3}$, which is ruled out by the bright thermal X-ray emission at the shock. If the density were this high, the shock would have rapidly become radiative and thermal X-ray emission would have declined significantly. Therefore, we show only the spectral model and data that we are able to fit in Figure \ref{fig:contours}.

Any synchrotron radiation from the relativistic electrons at the SNR shock interacting with the local shock-compressed magnetic field is hidden by the bright thermal X-ray emission from this remnant. Hence, we added a power-law component to the X-ray emission model to estimate an upper limit from the non-thermal component. The upper-limit on the flux is $F_X\leq5.4\times10^{-12}$\,ergs cm$^{-2}$ s$^{-1}$ for the range 0.2--10\,keV.

%It is very difficult to fit the gamma-ray data AND the radio+x-ray data with emission from the same electron particle distribution. The only way to make this happen is if the magnetic field is very much lower in this region than in the average ISM (if higher it violates the X-ray upperlimit).

With the current data, we can determine the inverse Compton (IC) decay maximum cut-off energy at 63.1\,TeV and a minimum at 12.6\,TeV. The corresponding spectral indices are 3.8 and 1.9, respectively. The maximum spectral index for the pion decay model is 3.4 with an upper limit on the cut-off energy of 316.2\,TeV (see Figure \ref{fig:contours}, right panel). Additionally in the pion decay model, we must impose a minimum spectral index of 2.5 (in momentum, or 1.5 in energy) because no acceleration process is believed to produce harder momentum (or energy) distributions than this. As a result, the minimum cut-off energy is unconstrained.

%This provides a minimum cut-off in energy to be 10\,TeV for inverse Compton emission and 20\,TeV for pion decay. 

Using the minimum spectral index of 2.5 in momentum, the minimum CR energy for electrons that could explain the observed $\gamma$-ray emission is $5.56\times10^{47}$\,erg\,($d$/6.3\,kpc)$^{2}$. The maximum CR energy for electrons is $1.74\times10^{49}$\,erg\,($d$/6.3\,kpc)$^{2}$, considering both distance estimates derived from \citet{dubner1993} and \citet{giacani2011} of 14\,kpc and 6.3\,kpc, respectively. The maximum CR electron energy corresponds to a spectral index of 4.2 in momentum and a cut-off of 200\,TeV. The total energy for CR protons, on the other hand, is better bound with a target density of $1$\,cm$^{-3}$. The minimum CR energy for protons is $4.0\times10^{50}$\,erg\,($n$/1\,cm$^{-3}$)($d$/6.3\,kpc)$^{2}$ and a minimum energy of $4.8\times10^{50}$\,erg\,(n/1\,cm$^{-3}$)($d$/6.3\,kpc)$^{2}$.

Based on the physical parameters of the best-fit SED models, both the leptonic and hadronic mechanisms could explain the observed $\gamma$-ray emission. If the GeV and TeV emission are generated from IC decay, it needs to either be unrelated to the SNR completely or the synchrotron population would need to be responsible for the IC $\gamma$-ray emission. However, as can be seen in Figure \ref{fig:xray}, there is no significant X-ray emission, thermal or nonthermal, that is detected beyond the SNR shell, suggesting no leptonic emission extending far beyond the SNR which poses a problem in explaining the large extension of HESS~J1702--420 and its emission peak occurring so far from the SNR. The other possibility would be two particle populations: one population of diffusing hadronic CRs that are escaping into the dense surroundings generating pion decay and the second population generating synchrotron radiation that is observed largely in radio. Because there is no hint of leptonic emission occurring outside of the SNR, the likely scenario is one where hadronic CRs are accelerated at the SNR forward shock where compact GeV emission is observed and escape into the surroundings where the CRs interact with a molecular cloud, generating the observed TeV emission. This is a plausible explanation since the SNR is known to be in a dense region of the Galaxy though no known molecular cloud (MC) at the position of the TeV emission shows a convincing correlation. If a molecular cloud exists and is being bombarded by the SNR CRs, the molecular cloud would rapidly thermalize and produce X-rays and, as previously mentioned, no thermal X-ray emission is detected beyond the SNR though this is not unusual if the interaction is relatively new. 
\\
\section{Conclusions}\label{sec:conclude}
The discovery and investigation of VHE $\gamma$-ray emission to the West of SNR~G344.7--0.1 is presented. Multi-wavelength data seems to point towards 2FHL~J1703.4--4145 originating from SNR  CRs accelerated by the forward shock that diffused into the ISM and interacted with a nearby molecular cloud, explaining the observed TeV emission, HESS~J1702--420. If this is the case, SNR~G344.7--0.1 would be a candidate for fresh CR acceleration. We perform and report a broadband spectral fitting and find that the $\gamma$-ray emission could be explained by either leptonic or hadronic scenarios, however, a pion decay scenario seems most likely based on the lack of leptonic emission seen beyond the SNR in the radio and X-ray. 

The presence of a large MC coincident with the observed $\gamma$-ray emission or other tracers of a SNR/MC interaction like those mentioned in Paper Ia (Sections 4.1 and 4.3) would be able to better determine the likelihood of the SNR freshly accelerating CRs. 
%The next catalog release from the HAWC observatory could help constrain the SED model, as HAWC can detect any VHE $\gamma$-ray emission from hundreds of GeV to $>100$\, TeV \citep[see e.g.][]{hawc2017, hawc2019}, however, no emission has been reported from HAWC to be associated or near SNR~G344.7--0.1 to date\footnote{Current public data sets from HAWC are available at \url{https://data.hawc-observatory.org/index.php}.} \citep{hawc2017}. 
 A deeper analysis in the VHE regime using instruments like \fermi, H.E.S.S., and the Cherenkov Telescope Array (CTA)\footnote{See \citet{cta2019}.} would ultimately improve our understanding of what emission mechanism is responsible for what is observed and thus the probability for freshly accelerated CRs in the SNR region.

%The discovery and investigation of a likely shock-cloud interaction taking place on the western edge of SNR~G344.7--0.1 is presented. Multi-wavelength data suggests the forward shock of the SNR is interacting with dense surrounding medium such as an HI cloud as a likely scenario. The data presented for 2FHL~J1703.4--4145 points towards the possibility of a site for CR acceleration. A broadband spectral fitting is reported for several particle populations that can explain the emission observed including leptonic and hadronic scenarios. {\color{red}Discuss which models are supported by data, implications, etc.}

%If the hadronic models prove to be most realistic in characterizing 2FHL~J0826.1$-$45.00 across available data, 2FHL~J0826.1$-$45.00 is not only a site of efficient particle acceleration but also poses as a candidate for fresh CR acceleration.

\section{Acknowledgements}
We acknowledge funding under NASA contract 80NSSC18K1716.
%CO data were obtained using the Mopra radio telescope, a part of the Australia Telescope National Facility which is funded by the Commonwealth of Australia for operation as a National Facility managed by CSIRO. The University of New South Wales (UNSW) digital filter bank (the UNSW-MOPS) used for the observations with Mopra was provided with support from the Australian Research Council (ARC), UNSW, Sydney and Monash Universities, as well as the CSIRO.

\bibliographystyle{aa}
\bibliography{apj_paper_layout}

\begin{thebibliography}{}
\expandafter\ifx\csname natexlab\endcsname\relax\def\natexlab#1{#1}\fi
\providecommand{\url}[1]{\href{#1}{#1}}
\providecommand{\dodoi}[1]{doi:~\href{http://doi.org/#1}{\nolinkurl{#1}}}
\providecommand{\doeprint}[1]{\href{http://ascl.net/#1}{\nolinkurl{http://ascl.net/#1}}}
\providecommand{\doarXiv}[1]{\href{https://arxiv.org/abs/#1}{\nolinkurl{https://arxiv.org/abs/#1}}}

\bibitem[{{Ackermann} {et~al.}(2016){Ackermann}, {Ajello}, {Atwood}, {Baldini},
  {Ballet}, {Barbiellini}, {Bastieri}, {Becerra Gonzalez}, {Bellazzini},
  {Bissaldi}, {Blandford}, {Bloom}, {Bonino}, {Bottacini}, {Brandt}, {Bregeon},
  {Bruel}, {Buehler}, {Buson}, {Caliandro}, {Cameron}, {Caputo}, {Caragiulo},
  {Caraveo}, {Cavazzuti}, {Cecchi}, {Charles}, {Chekhtman}, {Cheung}, {Chiang},
  {Chiaro}, {Ciprini}, {Cohen}, {Cohen-Tanugi}, {Cominsky}, {Conrad}, {Cuoco},
  {Cutini}, {D'Ammando}, {de Angelis}, {de Palma}, {Desiante}, {Di Mauro}, {Di
  Venere}, {Dom{\'{\i}}nguez}, {Drell}, {Favuzzi}, {Fegan}, {Ferrara}, {Focke},
  {Fortin}, {Franckowiak}, {Fukazawa}, {Funk}, {Furniss}, {Fusco}, {Gargano},
  {Gasparrini}, {Giglietto}, {Giommi}, {Giordano}, {Giroletti}, {Glanzman},
  {Godfrey}, {Grenier}, {Grondin}, {Guillemot}, {Guiriec}, {Harding}, {Hays},
  {Hewitt}, {Hill}, {Horan}, {Iafrate}, {Hartmann}, {Jogler},
  {J{\'o}hannesson}, {Johnson}, {Kamae}, {Kataoka}, {Kn{\"o}dlseder}, {Kuss},
  {La Mura}, {Larsson}, {Latronico}, {Lemoine-Goumard}, {Li}, {Li}, {Longo},
  {Loparco}, {Lott}, {Lovellette}, {Lubrano}, {Madejski}, {Maldera},
  {Manfreda}, {Mayer}, {Mazziotta}, {Michelson}, {Mirabal}, {Mitthumsiri},
  {Mizuno}, {Moiseev}, {Monzani}, {Morselli}, {Moskalenko}, {Murgia}, {Nuss},
  {Ohsugi}, {Omodei}, {Orienti}, {Orlando}, {Ormes}, {Paneque}, {Perkins},
  {Pesce-Rollins}, {Petrosian}, {Piron}, {Pivato}, {Porter}, {Rain{\`o}},
  {Rando}, {Razzano}, {Razzaque}, {Reimer}, {Reimer}, {Reposeur}, {Romani},
  {S{\'a}nchez-Conde}, {Saz Parkinson}, {Schmid}, {Schulz}, {Sgr{\`o}},
  {Siskind}, {Spada}, {Spandre}, {Spinelli}, {Suson}, {Tajima}, {Takahashi},
  {Takahashi}, {Takahashi}, {Thayer}, {Thompson}, {Tibaldo}, {Torres}, {Tosti},
  {Troja}, {Vianello}, {Wood}, {Wood}, {Yassine}, {Zaharijas}, \&
  {Zimmer}}]{ackermann2015}
{Ackermann}, M., {Ajello}, M., {Atwood}, W.~B., {et~al.} 2016, \apjs, 222, 5,
  \dodoi{10.3847/0067-0049/222/1/5}

\bibitem[{{Aharonian} {et~al.}(2006){Aharonian}, {Akhperjanian}, {Bazer-Bachi},
  {Beilicke}, {Benbow}, {Berge}, {Bernl{\"o}hr}, {Boisson}, {Bolz}, {Borrel},
  {Braun}, {Breitling}, {Brown}, {Chadwick}, {Chounet}, {Cornils},
  {Costamante}, {Degrange}, {Dickinson}, {Djannati-Ata{\"\i}}, {Drury},
  {Dubus}, {Emmanoulopoulos}, {Espigat}, {Feinstein}, {Fontaine}, {Fuchs},
  {Funk}, {Gallant}, {Giebels}, {Gillessen}, {Glicenstein}, {Goret},
  {Hadjichristidis}, {Hauser}, {Heinzelmann}, {Henri}, {Hermann}, {Hinton},
  {Hofmann}, {Holleran}, {Horns}, {Jacholkowska}, {de Jager}, {Kh{\'e}lifi},
  {Komin}, {Konopelko}, {Latham}, {Le Gallou}, {Lemi{\`e}re},
  {Lemoine-Goumard}, {Leroy}, {Lohse}, {Martin}, {Martineau-Huynh},
  {Marcowith}, {Masterson}, {McComb}, {de Naurois}, {Nolan}, {Noutsos},
  {Orford}, {Osborne}, {Ouchrif}, {Panter}, {Pelletier}, {Pita},
  {P{\"u}hlhofer}, {Punch}, {Raubenheimer}, {Raue}, {Raux}, {Rayner}, {Reimer},
  {Reimer}, {Ripken}, {Rob}, {Rolland}, {Rowell}, {Sahakian}, {Saug{\'e}},
  {Schlenker}, {Schlickeiser}, {Schuster}, {Schwanke}, {Siewert}, {Sol},
  {Spangler}, {Steenkamp}, {Stegmann}, {Tavernet}, {Terrier}, {Th{\'e}oret},
  {Tluczykont}, {Vasileiadis}, {Venter}, {Vincent}, {V{\"o}lk}, \&
  {Wagner}}]{aha2006}
{Aharonian}, F., {Akhperjanian}, A.~G., {Bazer-Bachi}, A.~R., {et~al.} 2006,
  \apj, 636, 777, \dodoi{10.1086/498013}

\bibitem[{{Ajello} {et~al.}(2017){Ajello}, {Atwood}, {Baldini}, {Ballet},
  {Barbiellini}, {Bastieri}, {Bellazzini}, {Bissaldi}, {Blandford}, {Bloom},
  {Bonino}, {Bregeon}, {Britto}, {Bruel}, {Buehler}, {Buson}, {Cameron},
  {Caputo}, {Caragiulo}, {Caraveo}, {Cavazzuti}, {Cecchi}, {Charles},
  {Chekhtman}, {Cheung}, {Chiaro}, {Ciprini}, {Cohen}, {Costantin}, {Costanza},
  {Cuoco}, {Cutini}, {D'Ammando}, {de Palma}, {Desiante}, {Digel}, {Di Lalla},
  {Di Mauro}, {Di Venere}, {Dom{\'\i}nguez}, {Drell}, {Dumora}, {Favuzzi},
  {Fegan}, {Ferrara}, {Fortin}, {Franckowiak}, {Fukazawa}, {Funk}, {Fusco},
  {Gargano}, {Gasparrini}, {Giglietto}, {Giommi}, {Giordano}, {Giroletti},
  {Glanzman}, {Green}, {Grenier}, {Grondin}, {Grove}, {Guillemot}, {Guiriec},
  {Harding}, {Hays}, {Hewitt}, {Horan}, {J{\'o}hannesson}, {Kensei}, {Kuss},
  {La Mura}, {Larsson}, {Latronico}, {Lemoine-Goumard}, {Li}, {Longo},
  {Loparco}, {Lott}, {Lubrano}, {Magill}, {Maldera}, {Manfreda}, {Mazziotta},
  {McEnery}, {Meyer}, {Michelson}, {Mirabal}, {Mitthumsiri}, {Mizuno},
  {Moiseev}, {Monzani}, {Morselli}, {Moskalenko}, {Negro}, {Nuss}, {Ohsugi},
  {Omodei}, {Orienti}, {Orlando}, {Palatiello}, {Paliya}, {Paneque}, {Perkins},
  {Persic}, {Pesce-Rollins}, {Piron}, {Porter}, {Principe}, {Rain{\`o}},
  {Rando}, {Razzano}, {Razzaque}, {Reimer}, {Reimer}, {Reposeur}, {Saz
  Parkinson}, {Sgr{\`o}}, {Simone}, {Siskind}, {Spada}, {Spandre}, {Spinelli},
  {Stawarz}, {Suson}, {Takahashi}, {Tak}, {Thayer}, {Thayer}, {Thompson},
  {Torres}, {Torresi}, {Troja}, {Vianello}, {Wood}, \& {Wood}}]{ajello2017}
{Ajello}, M., {Atwood}, W.~B., {Baldini}, L., {et~al.} 2017, \apjs, 232, 18,
  \dodoi{10.3847/1538-4365/aa8221}

\bibitem[{{Andersen} {et~al.}(2011){Andersen}, {Rho}, {Reach}, {Hewitt}, \&
  {Bernard}}]{spitzer2011}
{Andersen}, M., {Rho}, J., {Reach}, W.~T., {Hewitt}, J.~W., \& {Bernard}, J.~P.
  2011, \apj, 742, 7, \dodoi{10.1088/0004-637X/742/1/7}

\bibitem[{{Antonelli} {et~al.}(2009){Antonelli}, {Blasi}, {Bonanno},
  {Catalano}, {Covino}, {De Angelis}, {De Lotto}, {Ghigo}, {Ghisellini},
  {Israel}, {La Barbera}, {Pareschi}, {Persic}, {Roncadelli}, {Sacco},
  {Salvati}, {Tavecchio}, \& {Vallania}}]{antonelli2009}
{Antonelli}, L.~A., {Blasi}, P., {Bonanno}, G., {et~al.} 2009, ArXiv e-prints.
\newblock \doarXiv{0906.4114}

\bibitem[{{Arnaud}(1996)}]{phabs}
{Arnaud}, K.~A. 1996, Astronomical Society of the Pacific Conference Series,
  Vol. 101, {XSPEC: The First Ten Years}, ed. G.~H. {Jacoby} \& J.~{Barnes}, 17

\bibitem[{{Atwood} {et~al.}(2013){Atwood}, {Baldini}, {Bregeon}, {Bruel},
  {Chekhtman}, {Cohen-Tanugi}, {Drlica-Wagner}, {Granot}, {Longo}, {Omodei},
  {Pesce-Rollins}, {Razzaque}, {Rochester}, {Sgr{\`o}}, {Tinivella}, {Usher},
  \& {Zimmer}}]{atwood2013}
{Atwood}, W.~B., {Baldini}, L., {Bregeon}, J., {et~al.} 2013, \apj, 774, 76,
  \dodoi{10.1088/0004-637X/774/1/76}

\bibitem[{{Baring} {et~al.}(1999){Baring}, {Ellison}, {Reynolds}, {Grenier}, \&
  {Goret}}]{baring_1999}
{Baring}, M.~G., {Ellison}, D.~C., {Reynolds}, S.~P., {Grenier}, I.~A., \&
  {Goret}, P. 1999, \apj, 513, 311, \dodoi{10.1086/306829}

\bibitem[{{Bernl{\"o}hr} {et~al.}(2003){Bernl{\"o}hr}, {Carrol}, {Cornils},
  {Elfahem}, {Espigat}, {Gillessen}, {Heinzelmann}, {Hermann}, {Hofmann},
  {Horns}, {Jung}, {Kankanyan}, {Katona}, {Khelifi}, {Krawczynski}, {Panter},
  {Punch}, {Rayner}, {Rowell}, {Tluczykont}, \& {van Staa}}]{hess2003}
{Bernl{\"o}hr}, K., {Carrol}, O., {Cornils}, R., {et~al.} 2003, Astroparticle
  Physics, 20, 111, \dodoi{10.1016/S0927-6505(03)00171-3}

\bibitem[{{Braiding} {et~al.}(2018){Braiding}, {Wong}, {Maxted}, {Romano},
  {Burton}, {Blackwell}, {Filipovi{\'c}}, {Freeman}, {Indermuehle}, {Lau},
  {Rebolledo}, {Rowell}, {Snoswell}, {Tothill}, {Voisin}, \& {de
  Wilt}}]{mopra2018}
{Braiding}, C., {Wong}, G.~F., {Maxted}, N.~I., {et~al.} 2018, PASA, 35, e029,
  \dodoi{10.1017/pasa.2018.18}

\bibitem[{{Carrigan} {et~al.}(2013){Carrigan}, {Brun}, {Chaves}, {Deil},
  {Gast}, {Marandon}, \& {for the H.~E.~S.~S.~collaboration}}]{carrigan2013}
{Carrigan}, S., {Brun}, F., {Chaves}, R.~C.~G., {et~al.} 2013, ArXiv e-prints.
\newblock \doarXiv{1307.4868}

\bibitem[{{Castro} {et~al.}(2013{\natexlab{a}}){Castro}, {Lopez}, {Slane},
  {Yamaguchi}, {Ramirez-Ruiz}, \& {Figueroa-Feliciano}}]{castro_2013b}
{Castro}, D., {Lopez}, L.~A., {Slane}, P.~O., {et~al.} 2013{\natexlab{a}},
  \apj, 779, 49, \dodoi{10.1088/0004-637X/779/1/49}

\bibitem[{{Castro} {et~al.}(2013{\natexlab{b}}){Castro}, {Slane}, {Carlton}, \&
  {Figueroa-Feliciano}}]{castro_2013}
{Castro}, D., {Slane}, P., {Carlton}, A., \& {Figueroa-Feliciano}, E.
  2013{\natexlab{b}}, \apj, 774, 36, \dodoi{10.1088/0004-637X/774/1/36}

\bibitem[{{Chang} {et~al.}(2008){Chang}, {Konopelko}, \& {Cui}}]{chang2008}
{Chang}, C., {Konopelko}, A., \& {Cui}, W. 2008, \apj, 682, 1177,
  \dodoi{10.1086/589225}

\bibitem[{{Chawner} {et~al.}(2019){Chawner}, {Marsh}, {Matsuura}, {Gomez},
  {Cigan}, {De Looze}, {Barlow}, {Dunne}, {Noriega-Crespo}, \&
  {Rho}}]{chawner2019}
{Chawner}, H., {Marsh}, K., {Matsuura}, M., {et~al.} 2019, \mnras, 483, 70,
  \dodoi{10.1093/mnras/sty2942}

\bibitem[{{Chawner} {et~al.}(2020){Chawner}, {Gomez}, {Matsuura}, {Smith},
  {Papageorgiou}, {Rho}, {Noriega-Crespo}, {De Looze}, {Barlow}, {Cigan},
  {Dunne}, \& {Marsh}}]{chawner2020}
{Chawner}, H., {Gomez}, H.~L., {Matsuura}, M., {et~al.} 2020, \mnras, 493,
  2706, \dodoi{10.1093/mnras/staa221}

\bibitem[{{Cherenkov Telescope Array Consortium} {et~al.}(2019){Cherenkov
  Telescope Array Consortium}, {Acharya}, {Agudo}, {Al Samarai}, {Alfaro},
  {Alfaro}, {Alispach}, {Alves Batista}, {Amans}, {Amato}, {Ambrosi},
  {Antolini}, {Antonelli}, {Aramo}, {Araya}, {Armstrong}, {Arqueros},
  {Arrabito}, {Asano}, {Ashley}, {Backes}, {Balazs}, {Balbo}, {Ballester},
  {Ballet}, {Bamba}, {Barkov}, {Barres de Almeida}, {Barrio}, {Bastieri},
  {Becherini}, {Belfiore}, {Benbow}, {Berge}, {Bernardini}, {Bernardini},
  {Bernardos}, {Bernl{\"o}hr}, {Bertucci}, {Biasuzzi}, {Bigongiari}, {Biland},
  {Bissaldi}, {Biteau}, {Blanch}, {Blazek}, {Boisson}, {Bolmont}, {Bonanno},
  {Bonardi}, {Bonavolont{\`a}}, {Bonnoli}, {Bosnjak}, {B{\"o}ttcher},
  {Braiding}, {Bregeon}, {Brill}, {Brown}, {Brun}, {Brunetti}, {Buanes},
  {Buckley}, {Bugaev}, {B{\"u}hler}, {Bulgarelli}, {Bulik}, {Burton},
  {Burtovoi}, {Busetto}, {Canestrari}, {Capalbi}, {Capitanio}, {Caproni},
  {Caraveo}, {C{\'a}rdenas}, {Carlile}, {Carosi}, {Carqu{\'\i}n}, {Carr},
  {Casanova}, {Cascone}, {Catalani}, {Catalano}, {Cauz}, {Cerruti}, {Chadwick},
  {Chaty}, {Chaves}, {Chen}, {Chen}, {Chernyakova}, {Chikawa}, {Christov},
  {Chudoba}, {Cie{\'s}lar}, {Coco}, {Colafrancesco}, {Colin}, {Conforti},
  {Connaughton}, {Conrad}, {Contreras}, {Cortina}, {Costa}, {Costantini},
  {Cotter}, {Covino}, {Crocker}, {Cuadra}, {Cuevas}, {Cumani}, {D'A{\`\i}},
  {D'Ammando}, {D'Avanzo}, {D'Urso}, {Daniel}, {Davids}, {Dawson}, {Dazzi}, {De
  Angelis}, {de C{\'a}ssia dos Anjos}, {De Cesare}, {De Franco}, {de Gouveia
  Dal Pino}, {de la Calle}, {de los Reyes Lopez}, {De Lotto}, {De Luca}, {De
  Lucia}, {de Naurois}, {de O{\~n}a Wilhelmi}, {De Palma}, {De Persio}, {de
  Souza}, {Deil}, {Del Santo}, {Delgado}, {della Volpe}, {Di Girolamo}, {Di
  Pierro}, {Di Venere}, {D{\'\i}az}, {Dib}, {Diebold}, {Djannati-Ata{\"\i}},
  {Dom{\'\i}nguez}, {Dominis Prester}, {Dorner}, {Doro}, {Drass}, {Dravins},
  {Dubus}, {Dwarkadas}, {Ebr}, {Eckner}, {Egberts}, {Einecke}, {Ekoume},
  {Els{\"a}sser}, {Ernenwein}, {Espinoza}, {Evoli}, {Fairbairn},
  {Falceta-Goncalves}, {Falcone}, {Farnier}, {Fasola}, {Fedorova}, {Fegan},
  {Fernand ez-Alonso}, {Fern{\'a}ndez-Barral}, {Ferrand}, {Fesquet},
  {Filipovic}, {Fioretti}, {Fontaine}, {Fornasa}, {Fortson}, {Freixas
  Coromina}, {Fruck}, {Fujita}, {Fukazawa}, {Funk}, {F{\"u}{\ss}ling},
  {Gabici}, {Gadola}, {Gallant}, {Garcia}, {Garcia L{\'o}pez}, {Garczarczyk},
  {Gaskins}, {Gasparetto}, {Gaug}, {Gerard}, {Giavitto}, {Giglietto}, {Giommi},
  {Giordano}, {Giro}, {Giroletti}, {Giuliani}, {Glicenstein}, {Gnatyk},
  {Godinovic}, {Goldoni}, {G{\'o}mez-Vargas}, {Gonz{\'a}lez}, {Gonz{\'a}lez},
  {G{\"o}tz}, {Graham}, {Grand i}, {Granot}, {Green}, {Greenshaw}, {Griffiths},
  {Gunji}, {Hadasch}, {Hara}, {Hardcastle}, {Hassan}, {Hayashi}, {Hayashida},
  {Heller}, {Helo}, {Hermann}, {Hinton}, {Hnatyk}, {Hofmann}, {Holder},
  {Horan}, {H{\"o}randel}, {Horns}, {Horvath}, {Hovatta}, {Hrabovsky},
  {Hrupec}, {Humensky}, {H{\"u}tten}, {Iarlori}, {Inada}, {Inome}, {Inoue},
  {Inoue}, {Inoue}, {Iocco}, {Ioka}, {Iori}, {Ishio}, {Iwamura}, {Jamrozy},
  {Janecek}, {Jankowsky}, {Jean}, {Jung-Richardt}, {Jurysek}, {Kaaret},
  {Karkar}, {Katagiri}, {Katz}, {Kawanaka}, {Kazanas}, {Kh{\'e}lifi}, {Kieda},
  {Kimeswenger}, {Kimura}, {Kisaka}, {Knapp}, {Kn{\"o}dlseder}, {Koch},
  {Kohri}, {Komin}, {Kosack}, {Kraus}, {Krause}, {Krau{\ss}}, {Kubo}, {Kukec
  Mezek}, {Kuroda}, {Kushida}, {La Palombara}, {Lamanna}, {Lang}, {Lapington},
  {Le Blanc}, {Leach}, {Lees}, {Lefaucheur}, {Leigui de Oliveira}, {Lenain},
  {Lico}, {Limon}, {Lindfors}, {Lohse}, {Lombardi}, {Longo}, {L{\'o}pez},
  {L{\'o}pez-Coto}, {Lu}, {Lucarelli}, {Luque-Escamilla}, {Lyard}, {Maccarone},
  {Maier}, {Majumdar}, {Malaguti}, {Mandat}, {Maneva}, {Manganaro}, {Mangano},
  {Marcowith}, {Mar{\'\i}n}, {Markoff}, {Mart{\'\i}}, {Martin},
  {Mart{\'\i}nez}, {Mart{\'\i}nez}, {Masetti}, {Masuda}, {Maurin}, {Maxted},
  {Mazin}, {Medina}, {Melandri}, {Mereghetti}, {Meyer}, {Minaya}, {Mirabal},
  {Mirzoyan}, {Mitchell}, {Mizuno}, {Moderski}, {Mohammed}, {Mohrmann},
  {Montaruli}, {Moralejo}, {Morcuende-Parrilla}, {Mori}, {Morlino}, {Morris},
  {Morselli}, {Moulin}, {Mukherjee}, {Mundell}, {Murach}, {Muraishi}, {Murase},
  {Nagai}, {Nagataki}, {Nagayoshi}, {Naito}, {Nakamori}, {Nakamura}, {Niemiec},
  {Nieto}, {Niko{\l}ajuk}, {Nishijima}, {Noda}, {Nosek}, {Novosyadlyj},
  {Nozaki}, {O'Brien}, {Oakes}, {Ohira}, {Ohishi}, {Ohm}, {Okazaki}, {Okumura},
  {Ong}, {Orienti}, {Orito}, {Osborne}, {Ostrowski}, {Otte}, {Oya}, {Padovani},
  {Paizis}, {Palatiello}, {Palatka}, {Paoletti}, {Paredes}, {Pareschi},
  {Parsons}, {Pe'er}, {Pech}, {Pedaletti}, {Perri}, {Persic}, {Petrashyk},
  {Petrucci}, {Petruk}, {Peyaud}, {Pfeifer}, {Piano}, {Pisarski}, {Pita},
  {Pohl}, {Polo}, {Pozo}, {Prandini}, {Prast}, {Principe}, {Prokhorov},
  {Prokoph}, {Prouza}, {P{\"u}hlhofer}, {Punch}, {P{\"u}rckhauer}, {Queiroz},
  {Quirrenbach}, {Rain{\`o}}, {Razzaque}, {Reimer}, {Reimer}, {Reisenegger},
  {Renaud}, {Rezaeian}, {Rhode}, {Ribeiro}, {Rib{\'o}}, {Richtler}, {Rico},
  {Rieger}, {Riquelme}, {Rivoire}, {Rizi}, {Rodriguez}, {Rodriguez Fernandez},
  {Rodr{\'\i}guez V{\'a}zquez}, {Rojas}, {Romano}, {Romeo}, {Rosado}, {Rovero},
  {Rowell}, {Rudak}, {Rugliancich}, {Rulten}, {Sadeh}, {Safi-Harb}, {Saito},
  {Sakaki}, {Sakurai}, {Salina}, {S{\'a}nchez-Conde}, {Sandaker}, {Sandoval},
  {Sangiorgi}, {Sanguillon}, {Sano}, {Santand er}, {Sarkar}, {Satalecka},
  {Saturni}, {Schioppa}, {Schlenstedt}, {Schneider}, {Schoorlemmer},
  {Schovanek}, {Schulz}, {Schussler}, {Schwanke}, {Sciacca}, {Scuderi},
  {Seitenzahl}, {Semikoz}, {Sergijenko}, {Servillat}, {Shalchi}, {Shellard},
  {Sidoli}, {Siejkowski}, {Sillanp{\"a}{\"a}}, {Sironi}, {Sitarek}, {Sliusar},
  {Slowikowska}, {Sol}, {Stamerra}, {Stani{\v{c}}}, {Starling}, {Stawarz},
  {Stefanik}, {Stephan}, {Stolarczyk}, {Stratta}, {Straumann}, {Suomijarvi},
  {Supanitsky}, {Tagliaferri}, {Tajima}, {Tavani}, {Tavecchio}, {Tavernet},
  {Tayabaly}, {Tejedor}, {Temnikov}, {Terada}, {Terrier}, {Terzic}, {Teshima},
  {Testa}, {Thoudam}, {Tian}, {Tibaldo}, {Tluczykont}, {Todero Peixoto},
  {Tokanai}, {Tomastik}, {Tonev}, {Tornikoski}, {Torres}, {Torresi}, {Tosti},
  {Tothill}, {Tovmassian}, {Travnicek}, {Trichard}, {Trifoglio}, {Troyano
  Pujadas}, {Tsujimoto}, {Umana}, {Vagelli}, {Vagnetti}, {Valentino},
  {Vallania}, {Valore}, {van Eldik}, {Vand enbroucke}, {Varner}, {Vasileiadis},
  {Vassiliev}, {V{\'a}zquez Acosta}, {Vecchi}, {Vega}, {Vercellone}, {Veres},
  {Vergani}, {Verzi}, {Vettolani}, {Viana}, {Vigorito}, {Villanueva}, {Voelk},
  {Vollhardt}, {Vorobiov}, {Vrastil}, {Vuillaume}, {Wagner}, {Wagner},
  {Walter}, {Ward}, {Warren}, {Watson}, {Werner}, {White}, {White},
  {Wierzcholska}, {Wilcox}, {Will}, {Williams}, {Wischnewski}, {Wood},
  {Yamamoto}, {Yamazaki}, {Yanagita}, {Yang}, {Yoshida}, {Yoshiike},
  {Yoshikoshi}, {Zacharias}, {Zaharijas}, {Zampieri}, {Zand anel}, {Zanin},
  {Zavrtanik}, {Zavrtanik}, {Zdziarski}, {Zech}, {Zechlin}, {Zhdanov},
  {Ziegler}, \& {Zorn}}]{cta2019}
{Cherenkov Telescope Array Consortium}, {Acharya}, B.~S., {Agudo}, I., {et~al.}
  2019, {Science with the Cherenkov Telescope Array}, \dodoi{10.1142/10986}

\bibitem[{{Clark} {et~al.}(1975){Clark}, {Caswell}, \& {Green}}]{caswell1975}
{Clark}, D.~H., {Caswell}, J.~L., \& {Green}, A.~J. 1975, Australian Journal of
  Physics Astrophysical Supplement, 37, 1

\bibitem[{{Combi} {et~al.}(2010){Combi}, {Albacete Colombo},
  {L{\'o}pez-Santiago}, {Romero}, {S{\'a}nchez-Ayaso}, {Mart{\'\i}},
  {Luque-Escamilla}, {P{\'e}rez-Gonz{\'a}lez}, {Mu{\~n}oz-Arjonilla}, \&
  {S{\'a}nchez-Sutil}}]{combi2010}
{Combi}, J.~A., {Albacete Colombo}, J.~F., {L{\'o}pez-Santiago}, J., {et~al.}
  2010, \aap, 522, A50, \dodoi{10.1051/0004-6361/200913735}

\bibitem[{{Cui} {et~al.}(2016){Cui}, {P{\"u}hlhofer}, \&
  {Santangelo}}]{cui2016}
{Cui}, Y., {P{\"u}hlhofer}, G., \& {Santangelo}, A. 2016, \aap, 591, A68,
  \dodoi{10.1051/0004-6361/201628505}

\bibitem[{{Cui} {et~al.}(2019){Cui}, {Yang}, {He}, {Tam}, \&
  {Puhlhofer}}]{cui2019}
{Cui}, Y., {Yang}, R., {He}, X., {Tam}, P.~H.~T., \& {Puhlhofer}, G. 2019,
  arXiv e-prints, arXiv:1904.01761.
\newblock \doarXiv{1904.01761}

\bibitem[{{Drake} \& {Smale}(2016)}]{drake2016}
{Drake}, S.~A., \& {Smale}, A.~P. 2016, in AAS/High Energy Astrophysics
  Division, Vol.~15, AAS/High Energy Astrophysics Division \#15, 116.16

\bibitem[{{Dubner} {et~al.}(1993){Dubner}, {Moffett}, {Goss}, \&
  {Winkler}}]{dubner1993}
{Dubner}, G.~M., {Moffett}, D.~A., {Goss}, W.~M., \& {Winkler}, P.~F. 1993,
  \aj, 105, 2251, \dodoi{10.1086/116603}

\bibitem[{{Eagle} {et~al.}(2019){Eagle}, {Marchesi}, {Castro}, {Ajello},
  {Duvidovich}, \& {Tibaldo}}]{eagle2019}
{Eagle}, J., {Marchesi}, S., {Castro}, D., {et~al.} 2019, \apj, 870, 35,
  \dodoi{10.3847/1538-4357/aaf0ff}

\bibitem[{{Ferenc} \& {MAGIC Collaboration}(2005)}]{magic2005}
{Ferenc}, D., \& {MAGIC Collaboration}. 2005, Nuclear Instruments and Methods
  in Physics Research A, 553, 274, \dodoi{10.1016/j.nima.2005.08.085}

\bibitem[{{Fukushima} {et~al.}(2020){Fukushima}, {Yamaguchi}, {Slane}, {Park},
  {Katsuda}, {Sano}, {Lopez}, {Plucinsky}, {Kobayashi}, \&
  {Matsushita}}]{yama2020}
{Fukushima}, K., {Yamaguchi}, H., {Slane}, P.~O., {et~al.} 2020, arXiv
  e-prints, arXiv:2005.09664.
\newblock \doarXiv{2005.09664}

\bibitem[{{Funk}(2005)}]{funk2005}
{Funk}, S. 2005, International Cosmic Ray Conference, 4, 123

\bibitem[{{Gabici} {et~al.}(2007){Gabici}, {Aharonian}, \&
  {Blasi}}]{gabici2005}
{Gabici}, S., {Aharonian}, F.~A., \& {Blasi}, P. 2007, \apss, 309, 365,
  \dodoi{10.1007/s10509-007-9427-6}

\bibitem[{{Gabici} {et~al.}(2009){Gabici}, {Aharonian}, \&
  {Casanova}}]{gabici2009}
{Gabici}, S., {Aharonian}, F.~A., \& {Casanova}, S. 2009, \mnras, 396, 1629,
  \dodoi{10.1111/j.1365-2966.2009.14832.x}

\bibitem[{{Giacani} {et~al.}(2011){Giacani}, {Smith}, {Dubner}, \&
  {Loiseau}}]{giacani2011}
{Giacani}, E., {Smith}, M.~J.~S., {Dubner}, G., \& {Loiseau}, N. 2011, \aap,
  531, A138, \dodoi{10.1051/0004-6361/201116768}

\bibitem[{{Ginzburg} \& {Syrovatsky}(1965)}]{ginz1965}
{Ginzburg}, V.~L., \& {Syrovatsky}, S.~I. 1965, International Cosmic Ray
  Conference, 1, 53

\bibitem[{{H.~E.~S.~S. Collaboration} {et~al.}(2018){H.~E.~S.~S.
  Collaboration}, {Abdalla}, {Abramowski}, {Aharonian}, {Ait Benkhali},
  {Ang{\"u}ner}, {Arakawa}, {Arrieta}, {Aubert}, {Backes}, {Balzer}, {Barnard},
  {Becherini}, {Becker Tjus}, {Berge}, {Bernhard}, {Bernl{\"o}hr}, {Blackwell},
  {B{\"o}ttcher}, {Boisson}, {Bolmont}, {Bonnefoy}, {Bordas}, {Bregeon},
  {Brun}, {Brun}, {Bryan}, {B{\"u}chele}, {Bulik}, {Capasso}, {Carrigan},
  {Caroff}, {Carosi}, {Casanova}, {Cerruti}, {Chakraborty}, {Chaves}, {Chen},
  {Chevalier}, {Colafrancesco}, {Condon}, {Conrad}, {Davids}, {Decock}, {Deil},
  {Devin}, {deWilt}, {Dirson}, {Djannati-Ata{\"\i}}, {Domainko}, {Donath},
  {Drury}, {Dutson}, {Dyks}, {Edwards}, {Egberts}, {Eger}, {Emery},
  {Ernenwein}, {Eschbach}, {Farnier}, {Fegan}, {Fernand es}, {Fiasson},
  {Fontaine}, {F{\"o}rster}, {Funk}, {F{\"u}{\ss}ling}, {Gabici}, {Gallant},
  {Garrigoux}, {Gast}, {Gat{\'e}}, {Giavitto}, {Giebels}, {Glawion},
  {Glicenstein}, {Gottschall}, {Grondin}, {Hahn}, {Haupt}, {Hawkes},
  {Heinzelmann}, {Henri}, {Hermann}, {Hinton}, {Hofmann}, {Hoischen}, {Holch},
  {Holler}, {Horns}, {Ivascenko}, {Iwasaki}, {Jacholkowska}, {Jamrozy},
  {Jankowsky}, {Jankowsky}, {Jingo}, {Jouvin}, {Jung-Richardt}, {Kastendieck},
  {Katarzy{\'n}ski}, {Katsuragawa}, {Katz}, {Kerszberg}, {Khangulyan},
  {Kh{\'e}lifi}, {King}, {Klepser}, {Klochkov}, {Klu{\'z}niak}, {Komin},
  {Kosack}, {Krakau}, {Kraus}, {Kr{\"u}ger}, {Laffon}, {Lamanna}, {Lau},
  {Lees}, {Lefaucheur}, {Lemi{\`e}re}, {Lemoine-Goumard}, {Lenain}, {Leser},
  {Lohse}, {Lorentz}, {Liu}, {L{\'o}pez-Coto}, {Lypova}, {Marandon},
  {Malyshev}, {Marcowith}, {Mariaud}, {Marx}, {Maurin}, {Maxted}, {Mayer},
  {Meintjes}, {Meyer}, {Mitchell}, {Moderski}, {Mohamed}, {Mohrmann},
  {Mor{\r{a}}}, {Moulin}, {Murach}, {Nakashima}, {de Naurois}, {Ndiyavala},
  {Niederwanger}, {Niemiec}, {Oakes}, {O'Brien}, {Odaka}, {Ohm}, {Ostrowski},
  {Oya}, {Padovani}, {Panter}, {Parsons}, {Paz Arribas}, {Pekeur}, {Pelletier},
  {Perennes}, {Petrucci}, {Peyaud}, {Piel}, {Pita}, {Poireau}, {Poon},
  {Prokhorov}, {Prokoph}, {P{\"u}hlhofer}, {Punch}, {Quirrenbach}, {Raab},
  {Rauth}, {Reimer}, {Reimer}, {Renaud}, {de los Reyes}, {Rieger}, {Rinchiuso},
  {Romoli}, {Rowell}, {Rudak}, {Rulten}, {Safi-Harb}, {Sahakian}, {Saito},
  {Sanchez}, {Santangelo}, {Sasaki}, {Schand ri}, {Schlickeiser},
  {Sch{\"u}ssler}, {Schulz}, {Schwanke}, {Schwemmer}, {Seglar-Arroyo},
  {Settimo}, {Seyffert}, {Shafi}, {Shilon}, {Shiningayamwe}, {Simoni}, {Sol},
  {Spanier}, {Spir-Jacob}, {Stawarz}, {Steenkamp}, {Stegmann}, {Steppa},
  {Sushch}, {Takahashi}, {Tavernet}, {Tavernier}, {Taylor}, {Terrier},
  {Tibaldo}, {Tiziani}, {Tluczykont}, {Trichard}, {Tsirou}, {Tsuji}, {Tuffs},
  {Uchiyama}, {van der Walt}, {van Eldik}, {van Rensburg}, {van Soelen},
  {Vasileiadis}, {Veh}, {Venter}, {Viana}, {Vincent}, {Vink}, {Voisin},
  {V{\"o}lk}, {Vuillaume}, {Wadiasingh}, {Wagner}, {Wagner}, {Wagner}, {White},
  {Wierzcholska}, {Willmann}, {W{\"o}rnlein}, {Wouters}, {Yang}, {Zaborov},
  {Zacharias}, {Zanin}, {Zdziarski}, {Zech}, {Zefi}, {Ziegler}, {Zorn}, \&
  {{\.Z}ywucka}}]{hess2018}
{H.~E.~S.~S. Collaboration}, {Abdalla}, H., {Abramowski}, A., {et~al.} 2018,
  \aap, 612, A1, \dodoi{10.1051/0004-6361/201732098}

\bibitem[{{Hamilton} {et~al.}(1983){Hamilton}, {Sarazin}, \&
  {Chevalier}}]{chev1983}
{Hamilton}, A.~J.~S., {Sarazin}, C.~L., \& {Chevalier}, R.~A. 1983, \apjs, 51,
  115, \dodoi{10.1086/190841}

\bibitem[{{Holder} {et~al.}(2006){Holder}, {Atkins}, {Badran}, {Blaylock},
  {Bradbury}, {Buckley}, {Byrum}, {Carter-Lewis}, {Celik}, {Chow}, {Cogan},
  {Cui}, {Daniel}, {de la Calle Perez}, {Dowdall}, {Dowkontt}, {Duke},
  {Falcone}, {Fegan}, {Finley}, {Fortin}, {Fortson}, {Gibbs}, {Gillanders},
  {Glidewell}, {Grube}, {Gutierrez}, {Gyuk}, {Hall}, {Hanna}, {Hays}, {Horan},
  {Hughes}, {Humensky}, {Imran}, {Jung}, {Kaaret}, {Kenny}, {Kieda}, {Kildea},
  {Knapp}, {Krawczynski}, {Krennrich}, {Lang}, {LeBohec}, {Linton}, {Little},
  {Maier}, {Manseri}, {Milovanovic}, {Moriarty}, {Mukherjee}, {Ogden}, {Ong},
  {Petry}, {Perkins}, {Pizlo}, {Pohl}, {Quinn}, {Ragan}, {Reynolds}, {Roache},
  {Rose}, {Schroedter}, {Sembroski}, {Sleege}, {Steele}, {Swordy}, {Syson},
  {Toner}, {Valcarcel}, {Vassiliev}, {Wakely}, {Weekes}, {White}, {Williams},
  \& {Wagner}}]{ver2006}
{Holder}, J., {Atkins}, R.~W., {Badran}, H.~M., {et~al.} 2006, Astroparticle
  Physics, 25, 391, \dodoi{10.1016/j.astropartphys.2006.04.002}

\bibitem[{{Huang} \& {Thaddeus}(1985)}]{huang1985}
{Huang}, Y.~L., \& {Thaddeus}, P. 1985, \apj, 295, L13, \dodoi{10.1086/184528}

\bibitem[{{Kamae} {et~al.}(2006){Kamae}, {Karlsson}, {Mizuno}, {Abe}, \&
  {Koi}}]{kamae_2006}
{Kamae}, T., {Karlsson}, N., {Mizuno}, T., {Abe}, T., \& {Koi}, T. 2006, \apj,
  647, 692, \dodoi{10.1086/505189}

\bibitem[{{Kargaltsev} {et~al.}(2013){Kargaltsev}, {Rangelov}, \&
  {Pavlov}}]{karg2013}
{Kargaltsev}, O., {Rangelov}, B., \& {Pavlov}, G.~G. 2013, ArXiv e-prints.
\newblock \doarXiv{1305.2552}

\bibitem[{{Lau} {et~al.}(2019){Lau}, {Rowell}, {Voisin}, {Blackwell}, {Burton},
  {Braiding}, {Wong}, {Fukui}, \& {Casanova}}]{lau2019}
{Lau}, J.~C., {Rowell}, G., {Voisin}, F., {et~al.} 2019, \mnras, 483, 3659,
  \dodoi{10.1093/mnras/sty3326}

\bibitem[{{Mori}(2009)}]{mori_2009}
{Mori}, M. 2009, Astroparticle Physics, 31, 341,
  \dodoi{10.1016/j.astropartphys.2009.03.004}

\bibitem[{{Ong}(2014)}]{ong2014}
{Ong}, R.~A. 2014, Advances in Space Research, 53, 1483,
  \dodoi{10.1016/j.asr.2013.09.020}

\bibitem[{{Renaud}(2009)}]{renaud2009}
{Renaud}, M. 2009, ArXiv e-prints.
\newblock \doarXiv{0905.1287}

\bibitem[{{Sedov}(1959)}]{sedov1959}
{Sedov}, L.~I. 1959, {Similarity and Dimensional Methods in Mechanics}

\bibitem[{{Slane} {et~al.}(2015){Slane}, {Bykov}, {Ellison}, {Dubner}, \&
  {Castro}}]{castro2015}
{Slane}, P., {Bykov}, A., {Ellison}, D.~C., {Dubner}, G., \& {Castro}, D. 2015,
  \ssr, 188, 187, \dodoi{10.1007/s11214-014-0062-6}

\bibitem[{{Sudoh} {et~al.}(2019){Sudoh}, {Linden}, \& {Beacom}}]{halo2019}
{Sudoh}, T., {Linden}, T., \& {Beacom}, J.~F. 2019, arXiv e-prints,
  arXiv:1902.08203.
\newblock \doarXiv{1902.08203}

\bibitem[{{Taylor}(1950)}]{taylor1950}
{Taylor}, G. 1950, Proceedings of the Royal Society of London Series A, 201,
  159, \dodoi{10.1098/rspa.1950.0049}

\bibitem[{{Verner} {et~al.}(1996){Verner}, {Ferland}, {Korista}, \&
  {Yakovlev}}]{verner1996}
{Verner}, D.~A., {Ferland}, G.~J., {Korista}, K.~T., \& {Yakovlev}, D.~G. 1996,
  \apj, 465, 487, \dodoi{10.1086/177435}

\bibitem[{{Whiteoak} \& {Green}(1998)}]{whiteoak1996}
{Whiteoak}, J.~B.~Z., \& {Green}, A.~J. 1998, Astronomy Data Image Library

\bibitem[{{Wilms} {et~al.}(2000){Wilms}, {Allen}, \& {McCray}}]{wilms2000}
{Wilms}, J., {Allen}, A., \& {McCray}, R. 2000, \apj, 542, 914,
  \dodoi{10.1086/317016}

\bibitem[{{Yamaguchi} {et~al.}(2012){Yamaguchi}, {Tanaka}, {Maeda}, {Slane},
  {Foster}, {Smith}, {Katsuda}, \& {Yoshii}}]{yama2011}
{Yamaguchi}, H., {Tanaka}, M., {Maeda}, K., {et~al.} 2012, \apj, 749, 137,
  \dodoi{10.1088/0004-637X/749/2/137}

\bibitem[{{Yamauchi} {et~al.}(2005){Yamauchi}, {Ueno}, {Koyama}, \&
  {Bamba}}]{yama2005}
{Yamauchi}, S., {Ueno}, M., {Koyama}, K., \& {Bamba}, A. 2005, \pasj, 57, 459,
  \dodoi{10.1093/pasj/57.3.459}

\end{thebibliography}

\end{document}